\documentclass[conference]{IEEEtran}
\IEEEoverridecommandlockouts
\usepackage{titlesec}
\usepackage{tablefootnote}
\usepackage{cite}
\usepackage{amsmath,amssymb,amsfonts}
\usepackage{algorithm}
\usepackage{algpseudocode}
\usepackage{graphicx}
\usepackage{textcomp}
\usepackage{xcolor}
\usepackage{setspace}

\def\BibTeX{{\rm B\kern-.05em{\sc i\kern-.025em b}\kern-.08em
    T\kern-.1667em\lower.7ex\hbox{E}\kern-.125emX}}

\usepackage{bbm}
\usepackage{subcaption}
\usepackage{arydshln}
\usepackage[caption=false,font=footnotesize]{subfig}
\usepackage{comment}
\usepackage{multirow}

\usepackage{units}
\usepackage{etoolbox}
\usepackage[pagebackref=false,breaklinks=true,letterpaper=true,colorlinks=false,bookmarks=false]{hyperref}
\usepackage{array}
\usepackage{amsthm}

\newtheoremstyle{boldthm}
  {3pt}      
  {3pt}      
  {\itshape} 
  {}         
  {\bfseries} 
  {.}        
  {.5em}     
  {}         

\theoremstyle{boldthm}
\newtheorem{theorem}{Theorem}

\newtheoremstyle{proofoutline}
  {3pt}     
  {3pt}     
  {}        
  {}        
  {\bfseries} 
  {:}       
  {.5em}    
  {}        

\theoremstyle{proofoutline}

\newcommand{\calN}{{\cal N}}

\newcommand{\reals}{{\mathbb{R}}}

\DeclareMathOperator*{\argmax}{arg\,max}
\DeclareMathOperator*{\argmin}{arg\,min}

\title{Compute SNR-Optimal Analog-to-Digital Converters for Analog In-Memory Computing}

\author{
    Mihir Kavishwar and Naresh Shanbhag \\
    Department of Electrical and Computer Engineering, \\
    University of Illinois Urbana-Champaign, Urbana, IL 61801
}

\makeatletter
\providecommand\sf@counterlist{}  
\makeatother

\begin{document}

\maketitle

\begin{abstract}
Analog in-memory computing (AIMC) is an energy-efficient alternative to digital architectures for accelerating machine learning and signal processing workloads. However, its energy efficiency is limited by the high energy cost of the column analog-to-digital converters (ADCs). Reducing the ADC precision is an effective approach to lowering its energy cost. However, doing so also reduces the AIMC's computational accuracy thereby making it critical to identify the minimum precision required to meet a target accuracy. Prior works overestimate the ADC precision requirements by modeling quantization error as input-independent noise, maximizing the signal-to-quantization-noise ratio ($\mathrm{SQNR}$), and ignoring the discrete nature of ideal pre-ADC signal. We address these limitations by developing analytical expressions for estimating the compute signal-to-noise ratio ($\mathrm{CSNR}$), a true metric of accuracy for AIMCs, and propose CACTUS, an algorithm to obtain $\mathrm{CSNR}$-optimal ADC parameters. Using a circuit-aware behavioral model of an SRAM-based AIMC in a $\unit[28]{nm}$ CMOS process, we show that for a $256$-dimensional binary dot product, CACTUS reduces the ADC precision requirements by $\unit[3]{b}$ while achieving $\unit[6]{dB}$ higher $\mathrm{CSNR}$ over prior methods. We also delineate operating conditions under which our proposed $\mathrm{CSNR}$-optimal ADCs outperform conventional $\mathrm{SQNR}$-optimal ADCs.
\end{abstract}


\section{Introduction}

\begin{figure}[t]
    \centering
    \includegraphics[width=\linewidth]{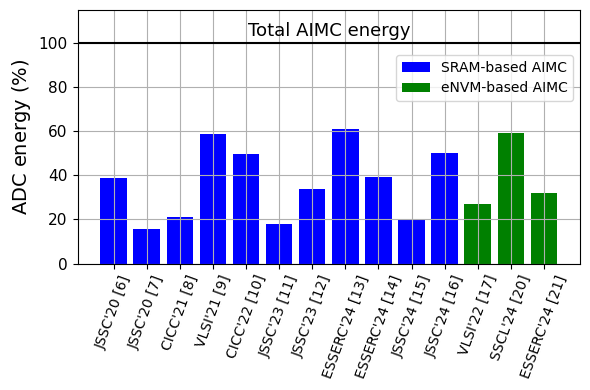}
     \caption{ADC energy contribution in state-of-the-art analog in-memory computing (AIMC) ICs.}
    \label{fig:adc_energy_contribution}
\end{figure}

Analog in-memory computing (AIMC) is emerging as a promising architecture for the energy-efficient execution of matrix-vector multiplication (MVM) operations, which dominate modern machine learning and signal processing workloads~\cite{imc_prospects, sebastian2020memory, imc_benchmarking, sun2023iccad}. AIMCs store large matrices in memory arrays, such as static random access memory (SRAM)~\cite{dong2020isscc, jia2020jssc, jiang2020jssc, papistas2021cicc, lee2021vlsi, choi2022cicc, sehgal2023jssc, wang2023jssc, lee2024esserc, choi2024esserc, yoshioka2024jssc, zhang2024jssc} or embedded non-volatile memory (eNVM)~\cite{correll2022vlsi, choi2023cicc, hung2023jssc, spetalnick2024sscl, yao2024esserc}. MVM operations are performed by applying input vectors to memory arrays, which compute dot products in the analog domain using charge or current accumulation, followed by digitization through analog-to-digital converters (ADCs). By integrating memory and computation, AIMCs eliminate frequent data transfers between separate memory and processing units - a major energy bottleneck in traditional von Neumann architectures~\cite{edvac, horowitz2014isscc}. However, this integration comes at the cost of significant ADC energy overhead. Fig.~\ref{fig:adc_energy_contribution} shows that ADCs can account for up to 60\% of the total energy consumption in state-of-the-art AIMCs today.

Lowering ADC precision reduces their energy cost~\cite{adc_survey}, thereby improving AIMC's energy efficiency. However, this also degrades AIMC's computational accuracy, creating an energy-accuracy trade-off. Therefore, there is much interest in determining the minimum ADC precision required in AIMCs to achieve a target computational accuracy, e.g.~\cite{gonugondla2020iccad, sakr2021tsp, murmann2021tvlsi, he2022iccad, sundar2024tcas2}. These works suffer from two deficiencies: 1) they estimate the AIMC's minimum ADC precision requirements by maximizing the ADCs signal-to-quantization-noise ratio ($\mathrm{SQNR}$) using it as a proxy for the computational accuracy of an AIMC; and 2) they model quantization errors as input-independent and the pre-ADC signal $V_\mathrm{pre-adc}$ as Gaussian distributed. Our work shows that both deficiencies lead to a significantly pessimistic estimate of the minimum required ADC precision. Since ADC energy quadruples in the noise-dominated regime~\cite{adc_survey,murmann2021tvlsi}, even a single bit reduction in ADC precision can be impactful in the overall energy efficiency of the AIMC.

\begin{figure*}[htb]
\setlength{\abovecaptionskip}{0pt} 
    \setlength{\belowcaptionskip}{5pt} 
    \centering    
    \begin{minipage}{0.53\linewidth}
    \begin{subfigure}{0.71\linewidth}
    \captionsetup{skip=0pt}
        \centering \includegraphics[trim = {0.1cm 0 0 1cm}, clip, width = \linewidth]{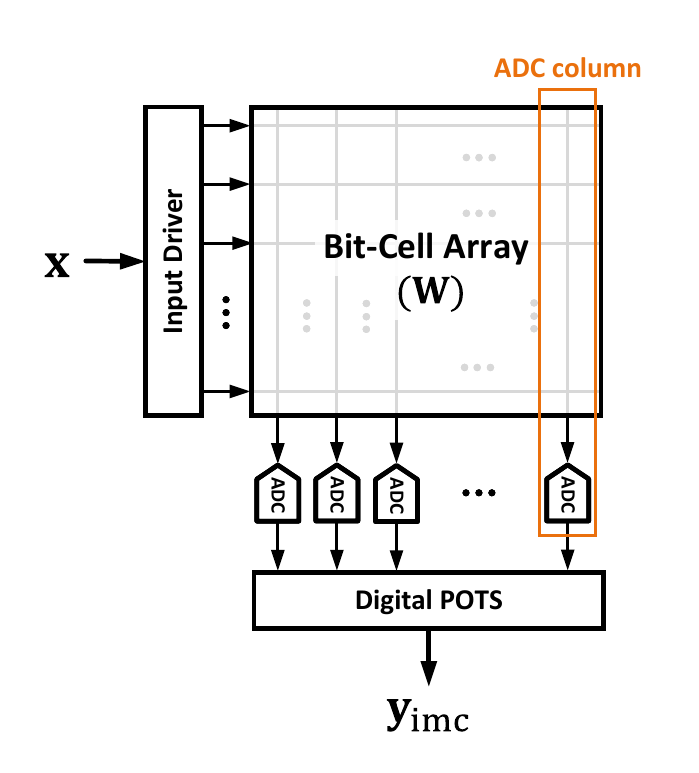}
        \caption{}
    \end{subfigure}
    \hfill
    \begin{subfigure}{0.26\linewidth}
    \captionsetup{skip=0pt}
        \centering        \includegraphics[trim = {0 0 0.1cm 0}, clip, width = \linewidth]{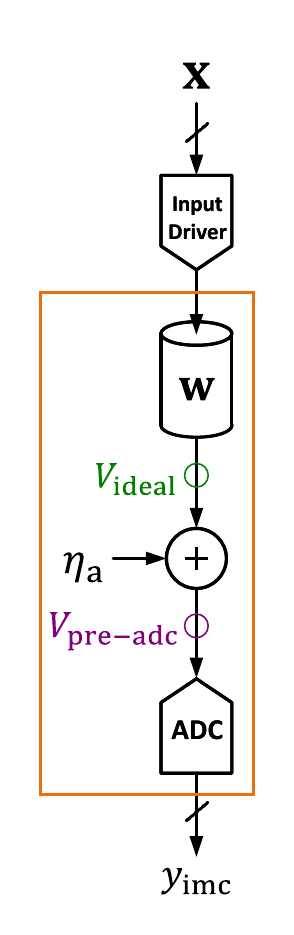}
        \caption{}
    \end{subfigure}  
    \end{minipage}
    \hfill
    \begin{minipage}{0.44\linewidth}
        \begin{subfigure}{\linewidth}
        \captionsetup{skip=0pt}
            \centering \includegraphics[width = \linewidth]{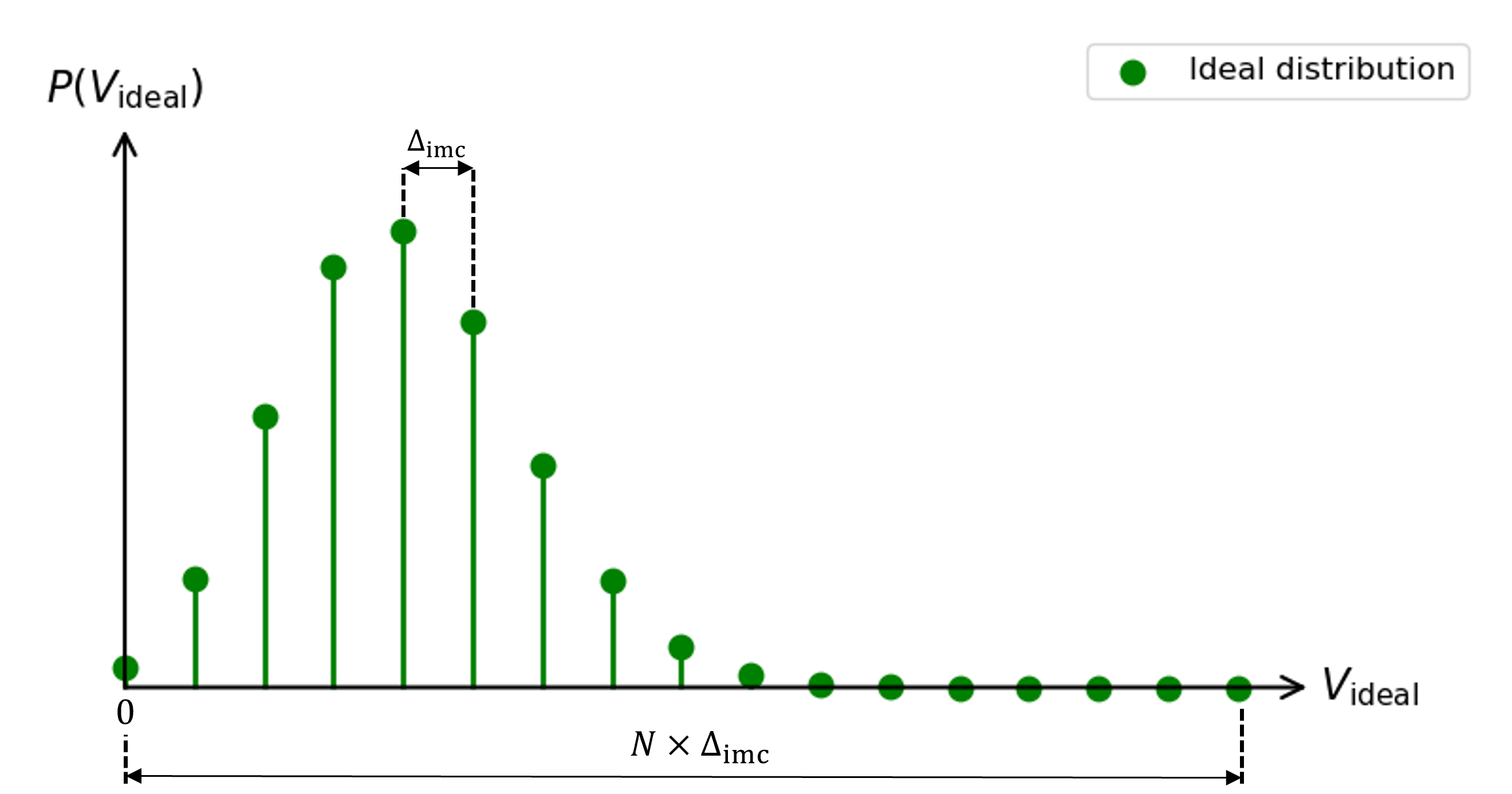}
        \caption{}
        \end{subfigure}
        \hfill
        \begin{subfigure}{\linewidth}
        \captionsetup{skip=0pt}
            \centering \includegraphics[width = \linewidth]{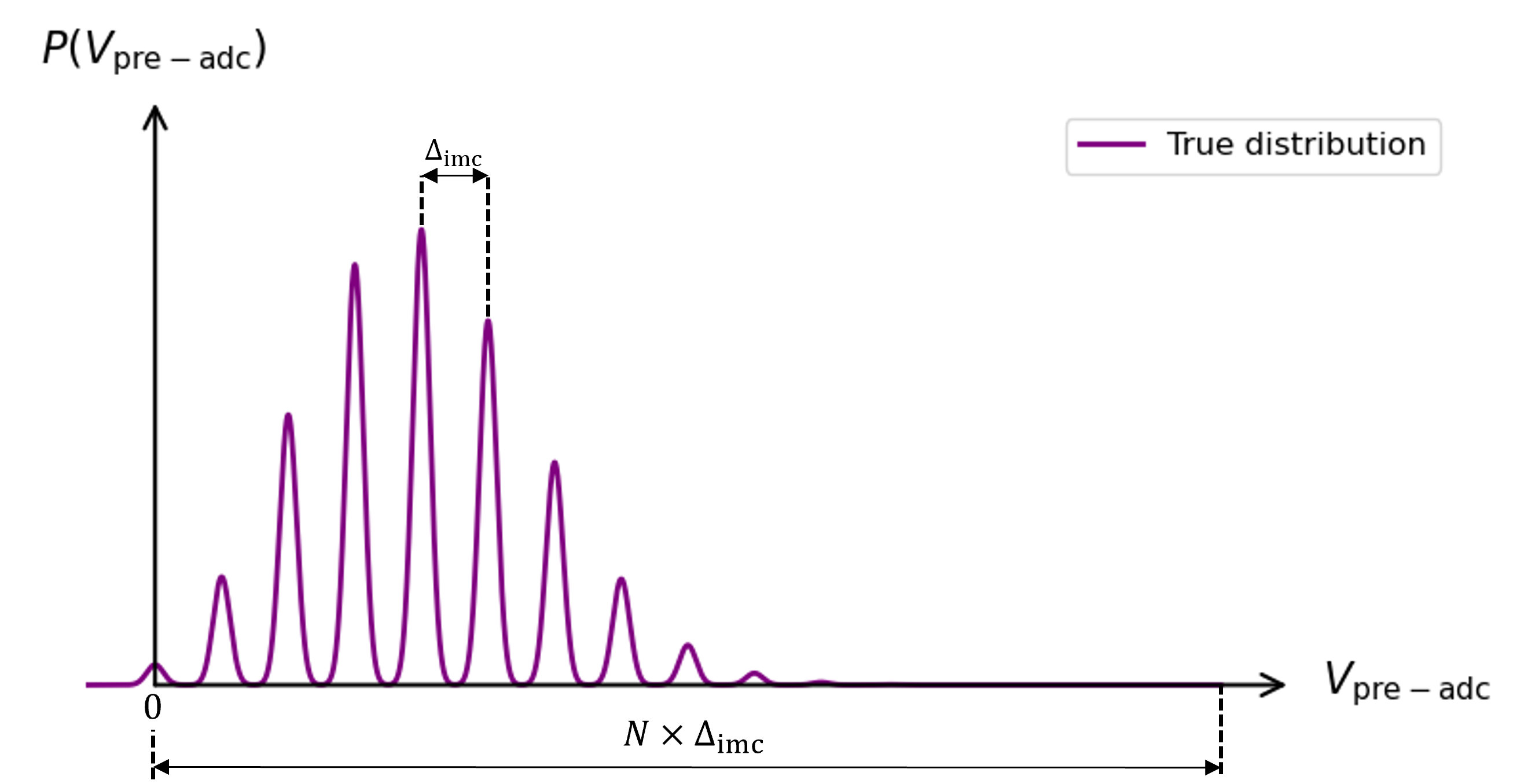}
        \caption{}
        \end{subfigure}
    \end{minipage}
    \caption{AIMC used for computing an MVM: (a) components of an AIMC bank, (b) model of a dot product computation in an ADC column showing the ideal voltage-domain dot product ($V_\mathrm{ideal}$), additive Gaussian noise ($\eta_\mathrm{a}$), and pre-ADC signal ($V_\mathrm{pre-adc}$). For a binary $N$-dimensional dot product: (c) $V_\mathrm{ideal}$ has a binomial distribution, and (d) $V_\mathrm{pre-adc}$ has a Gaussian mixture distribution with inter-level spacing of $\Delta_\mathrm{imc}$.}

    \label{fig:imc_dp_model}
\end{figure*}

In this paper, we determine the minimum ADC precision for AIMCs by: 1) maximizing the compute signal-to-noise ratio ($\mathrm{CSNR}$) instead of $\mathrm{SQNR}$. In doing so, we minimize the mean squared error between the ideal output of a digital dot product computation and that of the AIMC, i.e., the computational error, rather than the mean squared error between the input and output of an ADC as is done conventionally; and 2) we adopt a realistic AIMC model that assumes dependence between ADC quantization errors and its input, and treats the ADC input as a Gaussian mixture rather than a Gaussian. Fig.~\ref{fig:imc_dp_model} illustrates the difference between the distributions of the ideal voltage-domain dot product ($V_\mathrm{ideal}$) and the actual pre-ADC signal $V_\mathrm{pre-adc}$ on the bitlines. Note that $V_\mathrm{pre-adc}$ is a Gaussian mixture obtained by the addition of Gaussian noise representing ADC thermal noise and other noise sources, to a discrete-valued random variable $V_\mathrm{ideal}$ whose distribution tends to be binomial, e.g., $\mathrm{Bi}(N,0.25)$ for an $N$-dimension dot product computation. Prior works approximate this type of Gaussian mixture with a Gaussian which leads to pessimistic results.


\begin{figure*}[htb]
\setlength{\abovecaptionskip}{0pt} 
    \setlength{\belowcaptionskip}{5pt} 
    \centering    
    \begin{subfigure}{0.49\linewidth}
    \captionsetup{skip=0pt}
        \centering \includegraphics[trim = {0 0.1cm 0 0}, clip, width = \linewidth]{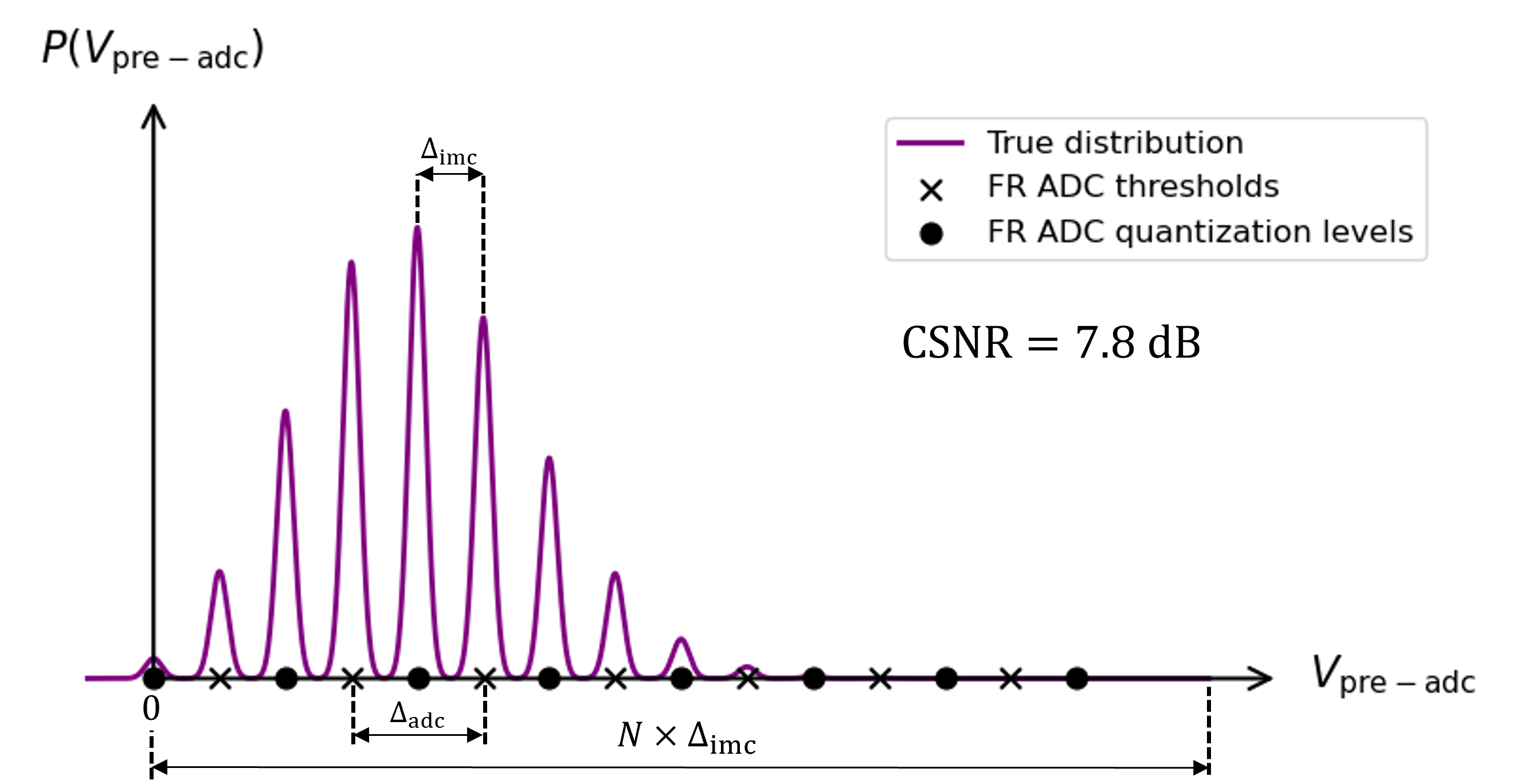}
        \caption{}
    \end{subfigure}
    \hfill
    \begin{subfigure}{0.49\linewidth}
    \captionsetup{skip=0pt}
        \centering \includegraphics[trim = {0 0.1cm 0 0}, clip, width = \linewidth]{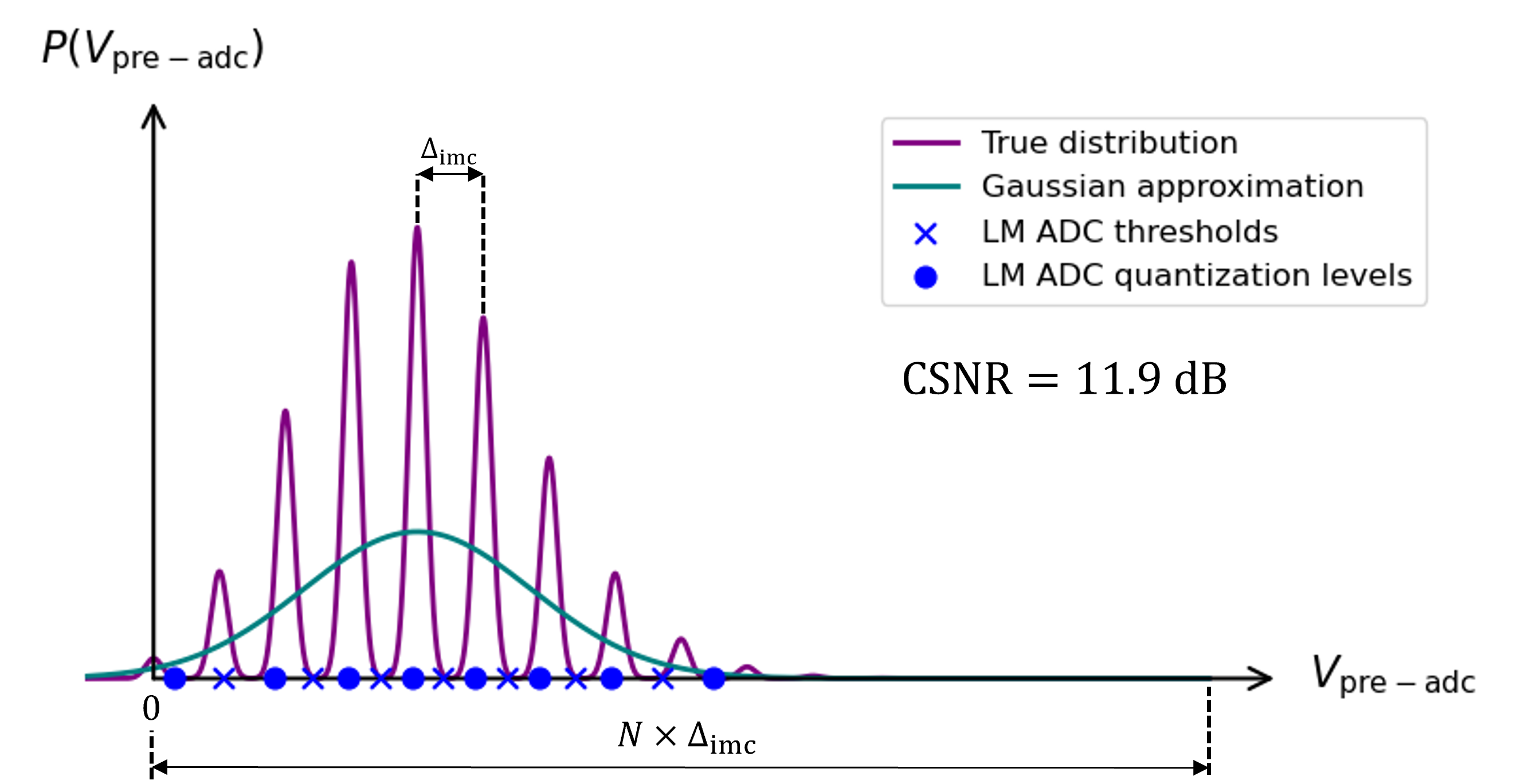}
        \caption{}
    \end{subfigure}
    \begin{subfigure}{0.49\linewidth}
    \captionsetup{skip=0pt}
        \centering \includegraphics[trim = {0 0.1cm 0 0}, clip, width = \linewidth]{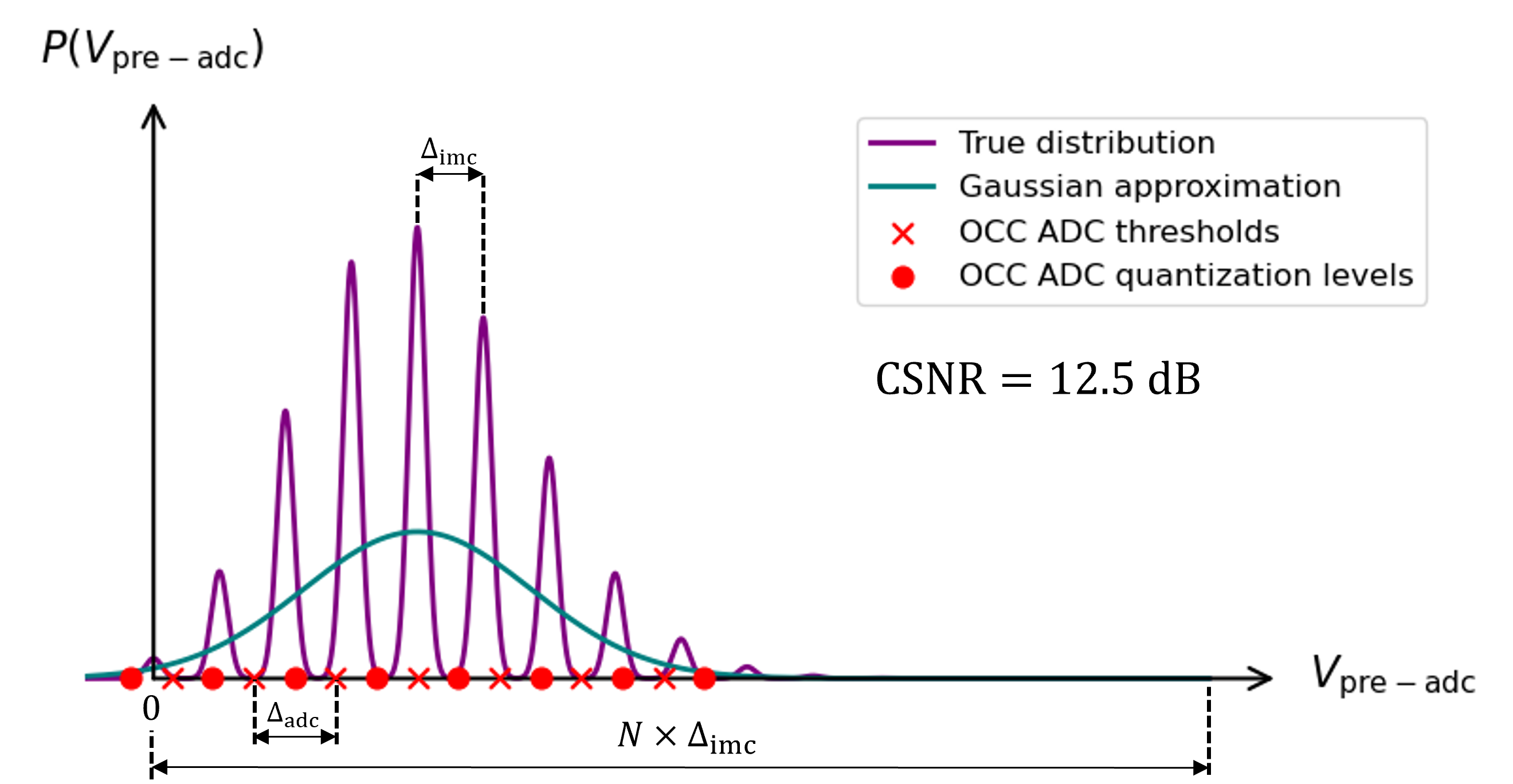}
        \caption{}
    \end{subfigure}
    \hfill
    \begin{subfigure}{0.49\linewidth}
    \captionsetup{skip=0pt}
        \centering \includegraphics[trim = {0 0.1cm 0 0}, clip, width = \linewidth]{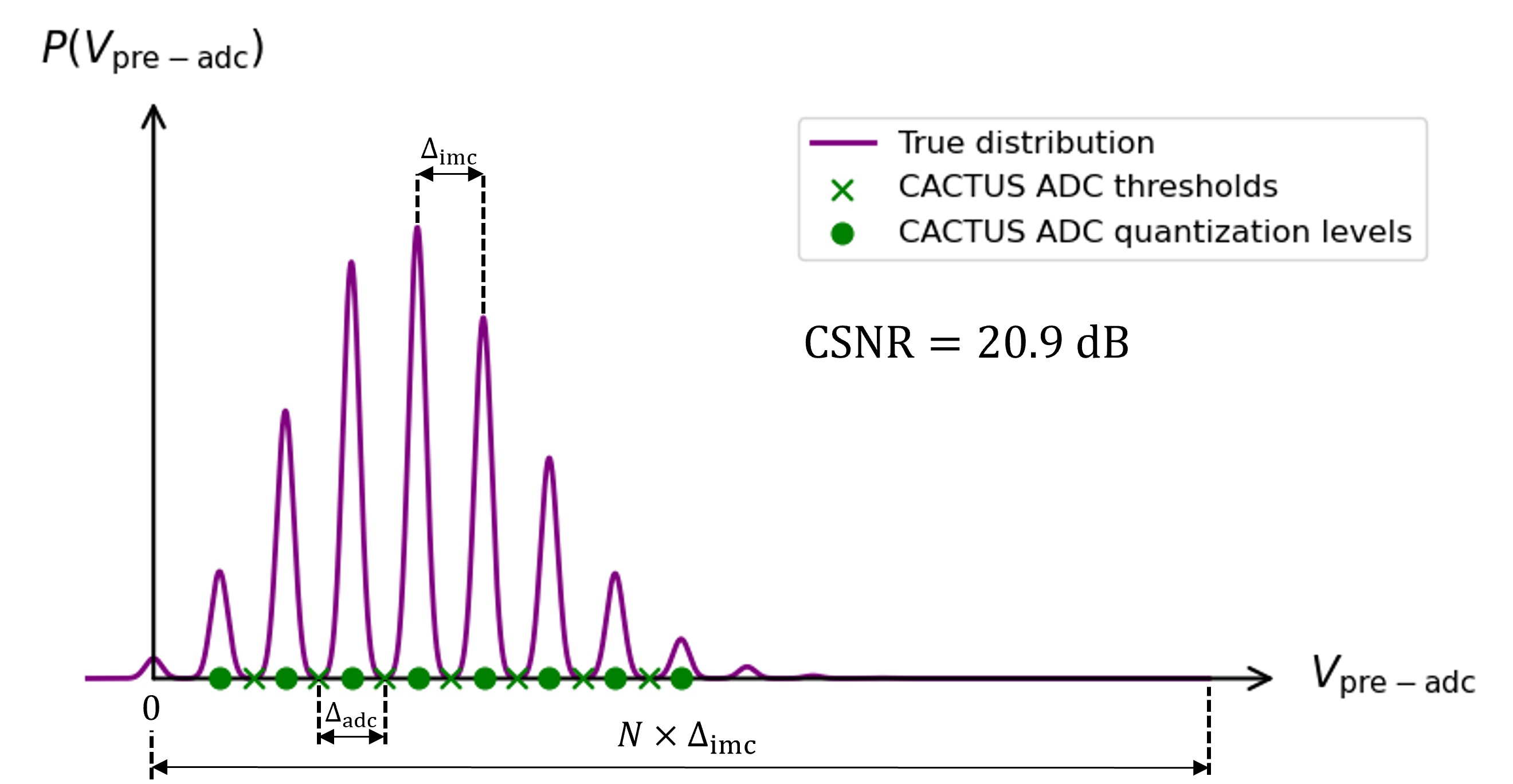}
        \caption{}
    \end{subfigure}
    \caption{Example of quantization thresholds of different types of column ADCs  for $N=16$, $\Delta_\mathrm{imc}=\unit[39.4]{mV}$, $\sigma_a =\unit[5]{mV}$, and $B_\mathrm{adc}=\unit[3]{b}$: (a) full range (FR), (b) Lloyd-Max (LM), (c) optimal clipping criteria (OCC), and the (d) proposed $\mathrm{CSNR}$-optimal ADC obtained from CACTUS. The $\mathrm{CSNR}$ values are obtained using Theorem~\ref{thm:main} and its extensions.}
    \label{fig:csnr_optimal_adc}
\end{figure*}

Our key contributions are summarized below:
\begin{itemize}
\item We propose $\mathrm{CSNR}$-optimal ADCs for AIMCs and highlight how they differ from $\mathrm{SQNR}$-optimal ADCs (Section~\ref{sec:main}). 
\item We derive analytical expressions for $\mathrm{CSNR}$ and $\mathrm{MSE_{dp}}$ in AIMCs, assuming a Gaussian mixture model for the ADC input and input-dependent quantization errors (Theorems~1 and~ 2 in Section~\ref{sec:main}).
\item Based on Theorems 1 and 2, we formulate CACTUS, an algorithm to identify $\mathrm{CSNR}$-optimal quantization thresholds given an ADC precision, and use it to determine the minimum ADC precision required to meet a target $\mathrm{CSNR}$ (Section~\ref{sec:main}(C)).
\item  We demonstrate, using a circuit-aware behavioral model of an SRAM-based AIMC in $\unit[28]{nm}$ CMOS, that CACTUS outperforms conventional baselines by achieving higher $\mathrm{CSNR}$ with lower ADC precision, e.g., for a $256$-dimensional binary dot product, CACTUS reduces the ADC precision requirements by $\unit[3]{b}$ while achieving $\unit[6]{dB}$ higher $\mathrm{CSNR}$ over prior methods.
\item  We delineate operating conditions under which our proposed $\mathrm{CSNR}$-optimal ADCs outperform conventional $\mathrm{SQNR}$-optimal ADCs.
\end{itemize}
A repository of our code can be found at: \href{https://github.com/mihirvk2/CSNR-optimal-ADC}{\textcolor{blue}{https://github.com/mihirvk2/CSNR-optimal-ADC}}. The rest of this paper is organized as follows. Section~\ref{sec:background} reviews the background on conventional ADCs and introduces $\mathrm{CSNR}$ as the accuracy metric for AIMC. Section~\ref{sec:main} presents a theoretical analysis of $\mathrm{CSNR}$-optimal ADCs and the CACTUS algorithm. Section~\ref{sec:results} discusses the simulation setup and results across various AIMC design parameters. Section~\ref{sec:conclusion} concludes the paper.

\section{Background and Related Works} \label{sec:background}

\subsection{Quantization Function in ADCs}

For a fixed ADC precision (\( B_\mathrm{adc} \)), the ADC is defined by its quantization function (\( Q_{\mathbf{t,r}} \)), which maps a continuous analog input to a discrete quantization level:  

\begin{equation}
    M = 2^{B_{\mathrm{adc}}} - 1
    \label{eq:num_qlevels}
\end{equation}
\begin{equation}
    Q_{\mathbf{t,r}}(V_\mathrm{in}) = 
    \begin{cases}
        r_0 &, \ \ \text{if } V_\mathrm{in} < t_1 \\
        r_k &, \ \ \text{if } t_k \leq V_\mathrm{in} < t_{k+1}, \  k \in \{1,2,\dots, M\} \\
        r_M &, \ \ \text{if } V_\mathrm{in} \geq t_M
    \end{cases}
    \label{eq:quantizer}
\end{equation}
where $V_\mathrm{in}$ represents the analog input to the ADC, \(Q_{\mathbf{t,r}}(\cdot) : \mathbb{R} \rightarrow \{r_0, r_1, \dots, r_M\}\) is the quantization function of the ADC with precision \(B_{\mathrm{adc}}\), thresholds \(\mathbf{t} = [t_1, \dots, t_M] \in \mathbb{R}^M\) with \(t_1 \leq \dots \leq t_M\), and quantization levels \(\mathbf{r} = [r_0, \dots, r_M] \in \mathbb{R}^{M+1}\) with $r_0 < t_1$, $t_k \leq r_k < t_{k+1} \ \forall \ k \in \{1,2,\dots, M-1\}$, $r_M \geq t_M$. The quantization levels \(\mathbf{r}\) are digitally encoded in hardware.

In uniform ADCs, the quantization function in~\eqref{eq:quantizer} is constrained to have uniformly spaced thresholds and quantization levels as follows:
\begin{equation}
    \Delta_\mathrm{adc} = \frac{t_M - t_1}{M-1}
    \label{eq:delta_adc}
\end{equation}
\begin{equation}
     t_k = t_1 + (k-1)\times \Delta_\mathrm{adc} \ \forall \ k \in \{1, 2, \dots , M \}
     \label{eq:uni_qlevels}
\end{equation}
\begin{equation}
     r_k = t_1 + (k-0.5)\times \Delta_\mathrm{adc} \ \forall  \ k \in \{0,1,\dots, M\}
    \label{eq:uni_qlevels}
\end{equation}
where \(\Delta_\mathrm{adc}\) is called the quantization step size of the ADC. The smallest and largest thresholds, \( t_1 \) and \( t_M \), are also known as the \emph{clipping thresholds}. Thus, quantization function of a uniform ADC can be completely expressed in terms of \( t_1 \) and \( t_M \). 

While in theory, non-uniform ADCs, i.e., ADCs with non-uniformly spaced quantization thresholds and levels are possible, such ADCs are difficult to design and their outputs cannot be processed using standard arithmetic blocks (adders and multipliers) due to the intrinsic non-linearity in the conversion process. Hence, in this paper, we focus on the commonly used uniform ADCs.

\subsection{$\mathrm{SQNR}$-optimal ADCs}
The quantization process in an ADC results in a quantization error/noise given by:
\begin{equation}
    q(V_\mathrm{in}) = Q_{\mathbf{t,r}}(V_\mathrm{in}) - V_\mathrm{in}
    \label{eq:q_err}
\end{equation}
where \(q(V_\mathrm{in})\) represents the quantization error corresponding to the input signal \(V_\mathrm{in}\). The ADC's accuracy is measured using the signal-to-quantization-noise ratio ($\mathrm{SQNR}$), defined as:
\begin{equation}
    \mathrm{SQNR}= \frac{\mathrm{Var}(V_\mathrm{in})}{\mathbb{E}[(Q_{\mathbf{t,r}}(V_\mathrm{in}) - V_\mathrm{in})^2]}  = \frac{\mathrm{Var}(V_\mathrm{in})}{\mathrm{MSE_{q}}} 
    \label{eq:sqnr_zero_mean}
\end{equation}
where \(V_\mathrm{in}\) is a random variable representing the analog ADC input, \(\mathrm{Var}(\cdot)\) denotes variance, $\mathbb{E}[\cdot]$ denotes expected value, and $\mathrm{MSE_q}$ is the mean squared quantization error. Thus, maximizing \(\mathrm{SQNR}\) is equivalent to minimizing $\mathrm{MSE_q}$. For a given ADC precision $B_\mathrm{adc}$, we define an \(\mathrm{SQNR}\)-optimal ADC as the solution to the following optimization problem:
\begin{equation}
\begin{split}
    \{\mathbf{t}^*, \mathbf{r}^*\} &= \argmax_{\mathbf{t}, \mathbf{r}} \ \mathrm{SQNR} = \argmin_{\mathbf{t}, \mathbf{r}} \ \mathrm{MSE_q}  \\
    &= \argmin_{\mathbf{t}, \mathbf{r}} \ \mathbb{E}[(Q_{\mathbf{t,r}}(V_\mathrm{in}) - V_\mathrm{in})^2]
\end{split}
\label{eq:msqe_adc}
\end{equation}
where \(\mathbf{t}^*\) and \(\mathbf{r}^*\) are the thresholds and quantization levels of the \(\mathrm{SQNR}\)-optimal ADC, respectively.

The Lloyd-Max (LM) algorithm~\cite{lloyd,max} is a well-established method for iteratively solving the unconstrained optimization problem in~\eqref{eq:msqe_adc} to obtain the $\mathrm{SQNR}$-optimal thresholds and quantization levels when the distribution of $V_\mathrm{in}$ is known. However, LM guarantees convergence only to a local minimum of $\mathrm{MSE_q}$, making it $\mathrm{SQNR}$-optimal only when the ADC input distribution has a single peak~\cite{LM_convergence}. Additionally, the convergence rate of LM can be extremely slow. Finally, LM-based ADCs are non-uniform creating several implementation challenges as indicated earlier.

The optimal clipping criterion (OCC)~\cite{sakr2021tsp} provides an implementation-friendly alternative to an LM quantizer. An OCC-based ADC employs uniformly-spaced quantization levels with the lowest ($t_1$) and highest ($t_M$) clipping levels chosen to maximize the $\mathrm{SQNR}$. In doing so, it achieves an $\mathrm{SQNR}$ close to that of the theoretically-optimal LM-quantizer while being easy to implement. However, OCC assumes a Gaussian distributed ADC input and input-independent quantization noise like the other works. 

In this paper, we will employ the full-range (FR),  LM- and OCC-based ADCs as baselines to compare our $\mathrm{CSNR}$-optimal ADCs against. Figure~\ref{fig:csnr_optimal_adc} shows the distinction between the three baselines and our proposed $\mathrm{CSNR}$-optimal ADCs. Note: the $\mathrm{CSNR}$-optimal ADC is also a uniform ADC. However, unlike the baselines, it accounts for the true ADC input distribution (Gaussian mixture) to set the quantization levels and thresholds. As a result, it can achieve a $\unit[8.4]{dB}$ higher $\mathrm{CSNR}$ than the best baseline in the example shown in Fig.~\ref{fig:csnr_optimal_adc}.

Our work is inspired by~\cite{BER_optimal_adc_1, BER_optimal_adc_2} where it was shown that ADCs that minimize the ultimate metric, bit-error rate ($\mathrm{BER}$), of a communication link require less precision than conventional $\mathrm{SQNR}$-optimal ADCs. However, unlike $\mathrm{BER}$-optimal ADCs, our $\mathrm{CSNR}$-optimal ADCs are uniform and therefore easy to implement in silicon, and target AIMCs. 

\subsection{$\mathrm{CSNR}$: Accuracy Metric for AIMCs} \label{subsec:accuracy}
Figure~\ref{fig:imc_dp_model}(a) shows the architecture of a standard AIMC bank comprising an input driver, bit-cell array (BCA), ADCs, and digital power-of-two-summation (POTS) logic to compute a dot product per ADC column as follows:
\begin{equation}
    y_\mathrm{ideal} = \mathbf{w}^\mathrm{T}\mathbf{x}
    \label{eq:ideal_dp}
\end{equation}
where \(\mathbf{x}\) is the input vector, \(\mathbf{w}\) is a column vector of the weights, 
and \(y_\mathrm{ideal}\) is the ideal dot product before POTS processing. Due to analog circuit non-idealities and quantization errors, the AIMC computation results in a computational error given by:
\begin{equation}
    e(\mathbf{w}, \mathbf{x}) = y_\mathrm{imc} - y_\mathrm{ideal}
    \label{eq:dp_err}
\end{equation}
where \(y_\mathrm{imc}\) is the AIMC-based dot product, and \(e(\mathbf{w}, \mathbf{x})\) denotes the dot product computational error corresponding to \(\mathbf{w}\) and \(\mathbf{x}\). The accuracy of the AIMC-based dot product computation is measured using the compute signal-to-noise ratio ($\mathrm{CSNR}$), defined as:
\begin{equation}
    \mathrm{CSNR} = \frac{\mathrm{Var}\left(y_\mathrm{ideal}\right)}{\mathrm{Var}\left( e(\mathbf{w}, \mathbf{x})\right)} = \frac{\mathrm{Var}\left(y_\mathrm{ideal}\right)}{\mathrm{Var}\left(y_\mathrm{imc} - y_\mathrm{ideal}\right)} 
    \label{eq:csndr}
\end{equation}
where \(\mathbf{w}\) and \(\mathbf{x}\) are random vectors representing the weight and input vectors, respectively, and \(y_\mathrm{ideal}\) and \(y_\mathrm{imc}\) are the random variables representing the ideal and AIMC-based dot product outputs, respectively. Since any constant offset in the AIMC-based dot product can be calibrated out digitally, we assume $e(\mathbf{w}, \mathbf{x})$ to be zero mean and rewrite~\eqref{eq:csndr} as:
\begin{equation}
    \mathrm{CSNR} = \frac{\mathrm{Var}\left(y_\mathrm{ideal}\right)}{\mathbb{E}\left[\left(y_\mathrm{imc} - y_\mathrm{ideal}\right)^2\right]} = \frac{\mathrm{Var}\left(y_\mathrm{ideal}\right)}{\mathrm{MSE_{dp}}} 
    \label{eq:csndr_zero_mean}
\end{equation}
where $\mathrm{MSE_{dp}}$ is the mean squared dot product error.

\(\mathrm{CSNR}\) is a task agnostic metric that quantifies the computational accuracy of AIMCs. In contrast, most prior works report only task specific metrics, such as classification accuracy in image classification tasks. However, such task specific metrics can make it difficult to compare different AIMC designs, as they are influenced by the algorithm mapped onto the AIMC. For instance, different machine learning models may yield different classification accuracies even when executed on the same AIMC. As a result, \(\mathrm{CSNR}\) is slowly gaining broader adoption~\cite{lee2024esserc, roy2024jxcdc, yoshioka2024jssc, roy2025jssc} as the standard bank level metric for evaluating AIMC's computational accuracy. Some works~\cite{roy2024jxcdc, roy2025jssc} refer to it as the compute signal-to-noise-plus-distortion ratio (\(\mathrm{CSNDR}\)) to highlight the distortion effects introduced by circuit parasitics.




\section{Proposed Compute SNR-optimal ADCs} \label{sec:main}
The AIMC-based dot product in~\eqref{eq:dp_err} can be expressed as:
\begin{equation}
    y_\mathrm{imc} = \mathrm{Dig}\left( Q_{\mathbf{t}, \mathbf{r}}\left(V_\mathrm{pre-adc} \right) \right)
    \label{eq:dig_yimc}
\end{equation}
where \(V_\mathrm{pre-adc}\) is the pre-ADC voltage-domain value of the AIMC dot product, \(Q_{\mathbf{t}, \mathbf{r}}(\cdot)\) is the quantization function defined in~\eqref{eq:quantizer}, and \(\mathrm{Dig}(\cdot)\) denotes the digital encoding function that maps the quantization levels to a digital representation (e.g., unsigned, two’s complement, etc.).  For a given $B_\mathrm{adc}$, we define a $\mathrm{CSNR}$-optimal ADC as the solution to the following optimization problem:
\begin{equation}
\begin{split}
    \{\mathbf{t^\dagger, r^\dagger}\} &= \argmax_{\mathbf{t}, \mathbf{r}} \ \mathrm{CSNR} = \argmin_{\mathbf{t}, \mathbf{r}} \ \mathrm{MSE}_\mathrm{dp} \\
    = \argmin_{\mathbf{t}, \mathbf{r}}  & \ \mathbb{E}\left[\left(\mathrm{Dig}\left( Q_{\mathbf{t}, \mathbf{r}}\left(V_\mathrm{pre-adc} \right) \right) - y_\mathrm{ideal}\right)^2\right] 
\end{split}
    \label{eq:msce_adc}
\end{equation}
where \(\mathbf{t}^\dagger\) and \(\mathbf{r}^\dagger\) are the thresholds and quantization levels of the \(\mathrm{CSNR}\)-optimal ADC, respectively.

\subsection{Modeling and Analysis of AIMCs}

To enable a direct mapping between the digital representation of dot product values and voltage-domain signals in AIMC, we assume \( y_\mathrm{ideal} \) is an unsigned discrete random variable with a known probability mass function:
\begin{equation}
     p_{y_\mathrm{ideal}}(y) = \Pr\{y_\mathrm{ideal} = y\}, \ \ y_\mathrm{ideal} \in \{0, 1, \dots, N\}
     \label{eq:ideal_pmf}
\end{equation}
where \( N \in \mathbb{N} \) is the maximum value of \( y_\mathrm{ideal} \). AIMCs computing binary $N$-dimensional dot products with Bernoulli distributed $\mathrm{Be}(0.5)$ weights and inputs, result in $y_\mathrm{ideal}\sim \mathrm{Bi}(N,0.25)$ with its voltage-domain equivalent given by:
\begin{equation}
    V_\mathrm{ideal} = y_\mathrm{ideal} \times \Delta_\mathrm{imc}
    \label{eq:v_ideal}
\end{equation}
where \( V_\mathrm{ideal} \) is a discrete random variable representing the ideal dot product in the voltage-domain, and \( \Delta_\mathrm{imc} \) is a fixed scaling factor. 

The pre-ADC voltage is influenced by various circuit non-idealities and can be modeled as\footnote{Embedded non-volatile memory (eNVM)-based AIMCs may require more complex models to account for distortion due to eNVM device non-idealities and wire parasitics~\cite{roy2024jxcdc}. However, most state-of-the-art SRAM-based AIMCs are well-approximated by~\eqref{eq:pre_adc}~\cite{gonugondla2020iccad, yoshioka2024jssc}.}: 
\begin{equation}
    V_\mathrm{pre-adc} = V_{\mathrm{ideal}} + \eta_\mathrm{a} \ \ , \ \ \eta_\mathrm{a} \sim \calN(0,\sigma^2_\mathrm{a})
\label{eq:pre_adc}
\end{equation}
where \(\eta_\mathrm{a}\) models thermal noise and charge injection. Fig.~\ref{fig:imc_dp_model} illustrates an example of distributions of $V_\mathrm{ideal}$ and $V_\mathrm{pre-adc}$ modeled as per~\eqref{eq:ideal_pmf}-\eqref{eq:pre_adc}. 

We chose the digital encoding function in~\eqref{eq:dig_yimc} such that when the voltage-domain quantization levels are scaled by $\Delta_\mathrm{imc}$ in~\eqref{eq:v_ideal}, the resulting fixed offset $\mu_\mathrm{off}$ in the computational error is calibrated out, i.e., $\mathbb{E}\left[ y_\mathrm{imc} - y_\mathrm{ideal}\right]=0$. The resulting $\mu_\mathrm{off}$ is given by:
\begin{align}
    \mu_\mathrm{off} = \mathbb{E}\left[ \frac{Q_{\mathbf{t,r}}\left(y_\mathrm{ideal}\times \Delta_\mathrm{imc} + \eta_\mathrm{a}\right) }{\Delta_\mathrm{imc}} - y_\mathrm{ideal} \right]\label{eq:zero_mean}
\end{align}

Thus, an AIMC is parametrized by: 1) $N$ dot product levels, 2) $p_{y_\mathrm{ideal}}$ distribution of ideal dot product, 3) $\Delta_\mathrm{imc}$ scaling factor, 4) $B_\mathrm{adc}$ ADC precision, 5) $\sigma_\mathrm{a}$ standard deviation of analog circuit noise, 6) $t_1$ smallest ADC clipping threshold, and 7) $t_M$ largest ADC clipping threshold. We denote this parameterization as: $\mathit{AIMC}\left(N, p_{y_\mathrm{ideal}}, \Delta_\mathrm{imc},  B_\mathrm{adc}, \sigma_\mathrm{a}, t_1, t_M \right)$.

We now derive analytical expressions for $\mu_\mathrm{off}$, \(\mathrm{MSE_{dp}}\) and \(\mathrm{CSNR}\) for AIMCs via Theorems 1 and 2. 

\begin{theorem}
For an AIMC\( \left(N, p_{y_\mathrm{ideal}}, \Delta_\mathrm{imc}, B_\mathrm{adc},\sigma_\mathrm{a}, t_1, t_M \right) \) modeled via~\eqref{eq:ideal_pmf}–\eqref{eq:pre_adc} with uniform ADCs modeled via~\eqref{eq:num_qlevels}–\eqref{eq:uni_qlevels},  the  $\mu_\mathrm{off}$ in~\eqref{eq:zero_mean} is given by:
\begin{equation}
    \begin{split}
    \mu_\mathrm{off} &= \frac{1}{\Delta_\mathrm{imc}} \sum_{y=0}^{N} p_{y_\mathrm{ideal}}(y) \Bigg\{ r_M - y\Delta_\mathrm{imc} \\
    & \ \ \ \ \ \ \ \ \ \ \ \ \ \ \ \ \ \ \ \ \ \ \ \ \ \ \ - \Delta_\mathrm{adc} \sum_{k=1}^{M} \Phi\left( \frac{t_{k} - y \Delta_\mathrm{imc}}{\sigma_\mathrm{a}} \right) \Bigg\}
\end{split}
\label{eq:thm_offset}
\end{equation}
where,
\begin{equation}
    \Phi(x) = \int_{-\infty}^x\frac{e^{-z^2/2}}{\sqrt{2\pi}} dz \ \forall \ x \in \reals
    \label{eq:normal_cdf}
\end{equation}
\label{thm:offset}
\end{theorem}
\noindent \textbf{Proof}: See Appendix.
\begin{theorem}
For an AIMC\( \left(N, p_{y_\mathrm{ideal}}, \Delta_\mathrm{imc}, B_\mathrm{adc},\sigma_\mathrm{a}, t_1, t_M \right) \) modeled via~\eqref{eq:ideal_pmf}–\eqref{eq:pre_adc} with uniform ADCs modeled via~\eqref{eq:num_qlevels}–\eqref{eq:uni_qlevels}, the $\mathrm{MSE}_\mathrm{dp}$ and $\mathrm{CSNR}$ in~\eqref{eq:csndr_zero_mean} are given by:
\begin{equation}
     \mathrm{MSE}_\mathrm{dp} = \alpha - \mu_\mathrm{off}^2
     \label{eq:mse_dp_soln}
\end{equation}
and
\begin{equation}
        \mathrm{CSNR} = \frac{\sum_{y= 0}^N y^2  p_\mathrm{y_{ideal}}(y) - \left( \sum_{y= 0}^N y p_\mathrm{y_{ideal}}(y) \right)^2 }{\alpha - \mu_\mathrm{off}^2}
        \label{eq:csnr_soln}
    \end{equation}
where,
\begin{equation}
\begin{split}
   \alpha =& \frac{1}{\Delta_\mathrm{imc}^2} \sum_{y=0}^{N} p_{y_\mathrm{ideal}}(y) \Bigg\{ \left( r_M - y\Delta_\mathrm{imc}\right)^2 \\
    & \ \ \ \ \ \ \ \ \ \  - 2\Delta_\mathrm{adc}   \sum_{k=1}^{M} \left(t_k - y\Delta_\mathrm{imc} \right) \Phi\left( \frac{t_{k} - y \Delta_\mathrm{imc}}{\sigma_\mathrm{a}} \right) \Bigg\} ,
\end{split}
\label{eq:msce_soln}
\end{equation}
$\mu_\mathrm{off}$ is defined in~\eqref{eq:thm_offset} and $\Phi(\cdot)$ is defined in~\eqref{eq:normal_cdf}.
    \label{thm:main}
\end{theorem}
\noindent \textbf{Proof}: See Appendix.


\subsection{Visualizing $\mathrm{CSNR}$ of AIMCs}

\begin{figure}[h]
\setlength{\abovecaptionskip}{0pt} 
    \setlength{\belowcaptionskip}{5pt} 
    \centering
    \captionsetup{skip=0pt}
        \includegraphics[trim = {0 0 0 0.2cm}, clip, width=0.95\linewidth]{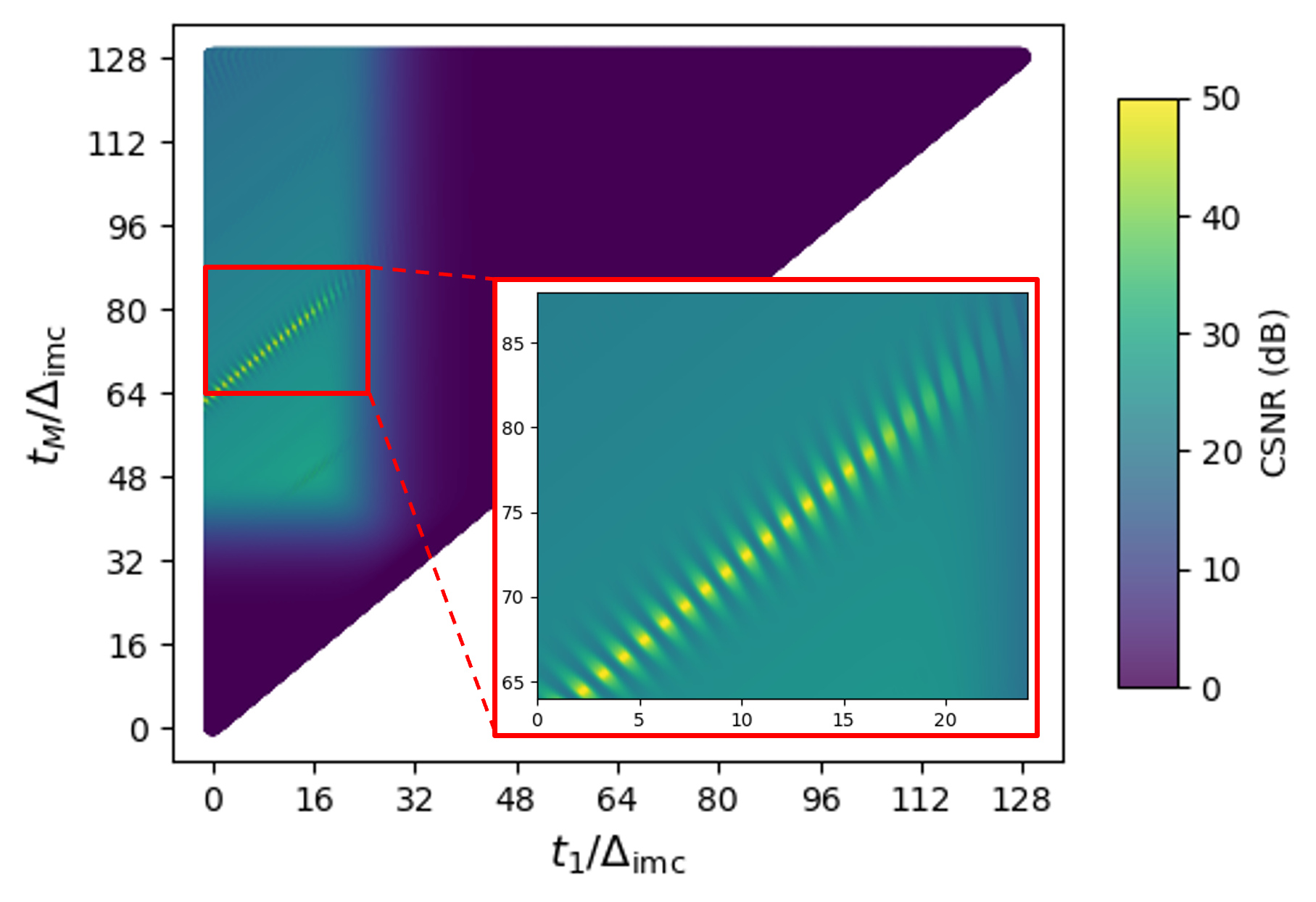}
    \caption{Theoretical compute signal-to-noise ratio ($\mathrm{CSNR}$) from Theorem~\ref{thm:main} as a function of ADC clipping thresholds, assuming $y_\mathrm{ideal}\sim \mathrm{Bi}(128,0.25),\  \Delta_\mathrm{imc} = 10\times \sigma_\mathrm{a}$, and $B_\mathrm{adc} = \unit[5]{b}$.}
    \label{fig:msce_thresholds_sweep}
\end{figure}

Based on an analysis of \( \mathrm{CSNR} \) across a range of AIMC configurations (see Fig.~\ref{fig:msce_thresholds_sweep} for a specific example), we make the following observations:
\begin{enumerate}
    \item $\mathrm{CSNR}$ is a non-convex function of $t_1$ and $t_M$, with multiple local maxima.
    \item $\mathrm{CSNR}$ exhibits large spikes at local maxima when $\Delta_\mathrm{imc} \gg \sigma_\mathrm{a}$.
    \item These spikes occur when $\Delta_\mathrm{adc}$ becomes an integer multiple of $\Delta_\mathrm{imc}$ and the ADC quantization levels align with the high-probability levels of $y_\mathrm{ideal}$.
\end{enumerate}

Based on these observations, we propose an algorithm, CACTUS, to solve~\eqref{eq:msce_adc} under the constraint of uniform quantization.

\subsection{CACTUS Algorithm}

\begin{algorithm}[h]
\caption{\textbf{C}SNR-optimal \textbf{A}DC \textbf{C}lipping \textbf{T}hresholds \textbf{S}earch (CACTUS)}
\label{alg:cactus}
\begin{spacing}{1.2} 
\begin{algorithmic}[1]
\State \textbf{Inputs:} $N$, $p_{y_\mathrm{ideal}}$, $\Delta_{\mathrm{imc}}$, $B_{\mathrm{adc}}$, $\sigma_{\mathrm{a}}$
\State \textbf{Outputs: $t_1^\dagger$, $t_M^\dagger, \mathrm{MSE}_\mathrm{dp}^{\min}$,  $\mathrm{CSNR}^{\max}$} 
\If{$B_{\mathrm{adc}} \geq \log_2 N$}
    \State $t_1^\dagger \gets 0.5  \Delta_{\mathrm{imc}}$ , $t_M^\dagger \gets (M-0.5) \Delta_{\mathrm{imc}}$
    \State $\mathrm{MSE}_\mathrm{dp}^{\min}$,$\mathrm{CSNR}^{\max} \gets$ \textcolor{red}{{Th}.~\ref{thm:main}} with parameters:
    \hspace*{9em}$(N, p_{y_\mathrm{ideal}}, \Delta_{\mathrm{imc}}, B_{\mathrm{adc}},\sigma_{\mathrm{a}}, t_1^\dagger, t_M^\dagger)$
    \State \Return $t_1^\dagger, t_M^\dagger, \mathrm{MSE}_\mathrm{dp}^{\min}, \mathrm{CSNR}^{\min}$ 
\Else
\State   $t_1^\dagger, t_M^\dagger \gets 0$, $\mathrm{MSE}_\mathrm{dp}^\mathrm{min} \gets \infty$, $\mathrm{CSNR}^\mathrm{max} \gets -\infty$, \hspace*{1.3em} $M \gets 2^{B_{\mathrm{adc}}} - 1$,  $k \gets 1$
    \While{$(M - 0.5) k  <   N$}
        \State $\Delta_{\mathrm{adc}} \gets k  \Delta_{\mathrm{imc}}$, $l \gets 0$
        \While{$(M-1)k+ l+0.5 <  N$}
            \State $t_1' \gets (l+0.5) \Delta_{\mathrm{imc}}$, $t_M' \gets t_1' + (M-1) \Delta_{\mathrm{adc}}$
            \State $\mathrm{MSE}_\mathrm{dp}' , \mathrm{CSNR}'\gets$ \textcolor{red}{{Th}.~\ref{thm:main}} with parameters:
            \hspace*{9em}$(N, p_{y_\mathrm{ideal}}, \Delta_{\mathrm{imc}}, B_{\mathrm{adc}},\sigma_{\mathrm{a}}, t_1', t_M')$
            \If{$\mathrm{MSE}_\mathrm{dp}' < \mathrm{MSE}_\mathrm{dp}^{\min}$}
               \State $t_1^\dagger , t_M^\dagger \gets t_1', t_M'$
                \State $\mathrm{MSE}_\mathrm{dp}^{\min}, \mathrm{CSNR}^{\max} \gets \mathrm{MSE}_\mathrm{dp}', \mathrm{CSNR}'$
            \EndIf
            \State $l \gets l + 1$
        \EndWhile
        \State $k \gets k + 1$
    \EndWhile
    \State \Return $t_1^\dagger$,$ t_M^\dagger$, $\mathrm{MSE}_\mathrm{dp}^{\min}$, $\mathrm{CSNR}^{\max}$ 
\EndIf
\end{algorithmic}
\end{spacing}
\end{algorithm}

Algorithm~\ref{alg:cactus} (CACTUS) estimates the CSNR-optimal clipping thresholds for uniform ADCs in AIMCs, for a given ADC precision, and provides estimates of the minimum achievable \(\mathrm{MSE}_\mathrm{dp}\) and the corresponding maximum \(\mathrm{CSNR}\). CACTUS first checks whether the ADC precision is sufficient to represent all \(N\) possible dot product values (i.e., if \(B_{\mathrm{adc}} \geq \log_2 N\)). If so, it directly sets the clipping thresholds such that the quantization levels match the $V_\mathrm{ideal}$ levels, and estimates \(\mathrm{MSE}_\mathrm{dp}\) and \(\mathrm{CSNR}\) using Theorem~\ref{thm:main}. Otherwise, CACTUS performs an iterative search over a finite set of candidate threshold pairs that satisfy two conditions: 1) \(\Delta_{\mathrm{adc}}\) is an integer multiple of \(\Delta_{\mathrm{imc}}\), and 2) each threshold lies exactly in between its two nearest \(V_\mathrm{ideal}\) levels. For each candidate \((t_1', t_M')\), CACTUS estimates \(\mathrm{MSE}_\mathrm{dp}\) and \(\mathrm{CSNR}\) using Theorem~\ref{thm:main}, and selects the candidate with the lowest \(\mathrm{MSE}_\mathrm{dp}\).

\begin{algorithm}
    \caption{Minimum ADC precision requirement}
    \label{alg:mpr}
    \begin{spacing}{1.2} 
    \begin{algorithmic}[1]
    \State \textbf{Inputs:} $N$, $p_{y_\mathrm{ideal}}$, $\Delta_{\mathrm{imc}}$, $\sigma_{\mathrm{a}}$, $\mathrm{CSNR^{spec}}$
    \State \textbf{Output:} $B_\mathrm{adc}^\mathrm{min}$
    \State $B_\mathrm{adc}^\mathrm{min} \gets 1$
    \While{$B_\mathrm{adc}^\mathrm{min} \leq \lceil \log_2 N \rceil$}
        \State $\mathrm{CSNR}^{\max} \gets$ \textcolor{red}{CACTUS} $(N, p_{y_\mathrm{ideal}}, \Delta_{\mathrm{imc}}, B_\mathrm{adc}^\mathrm{min}, \sigma_{\mathrm{a}})$
        \If{$\mathrm{CSNR}^{\max} \geq \mathrm{CSNR}^{\mathrm{spec}}$}
            \State \Return $B_\mathrm{adc}^\mathrm{min}$
        \Else
            \State $B_\mathrm{adc}^\mathrm{min} \gets B_\mathrm{adc}^\mathrm{min} + 1$
        \EndIf
    \EndWhile
    \end{algorithmic}
    \end{spacing}
\end{algorithm}
Algorithm~\ref{alg:mpr} determines the minimum ADC precision required to meet a target CSNR specification. At each iteration, the algorithm estimates the maximum achievable \(\mathrm{CSNR}\) using CACTUS and increments the ADC precision \(B_\mathrm{adc}^\mathrm{min}\) until the target \(\mathrm{CSNR^{spec}}\) is met.

\section{Simulation Results} \label{sec:results}

\subsection{Simulation Setup}

\begin{figure}[h]
    \centering
    \includegraphics[width=\linewidth]{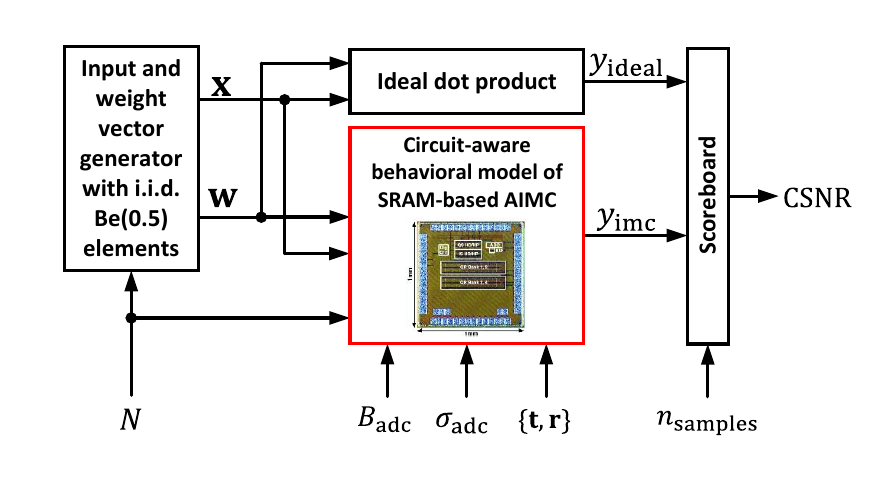}
    \caption{Simulation setup for evaluating $\mathrm{CSNR}$ of an SRAM-based AIMC computing $y_\mathrm{ideal} = \mathbf{w}^\mathrm{T}\mathbf{x}$.  }
    \label{fig:sim_setup}
\end{figure}

\begin{figure*}[tb]
\setlength{\abovecaptionskip}{0pt} 
    \setlength{\belowcaptionskip}{5pt} 
    \centering    
    \begin{subfigure}{0.32\linewidth}
    \captionsetup{skip=0pt}
        \centering        \includegraphics[width = \linewidth]{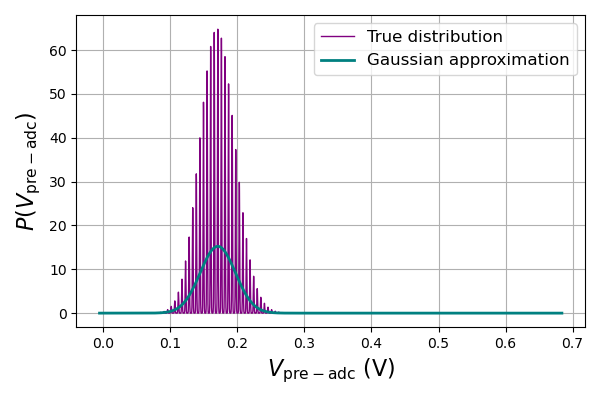}
        \caption{$N = 128$, $\sigma_\mathrm{adc} = \unit[0.5]{mV}$ }
    \end{subfigure}
    \hfill
    \begin{subfigure}{0.32\linewidth}
    \captionsetup{skip=0pt}
        \centering        \includegraphics[width = \linewidth]{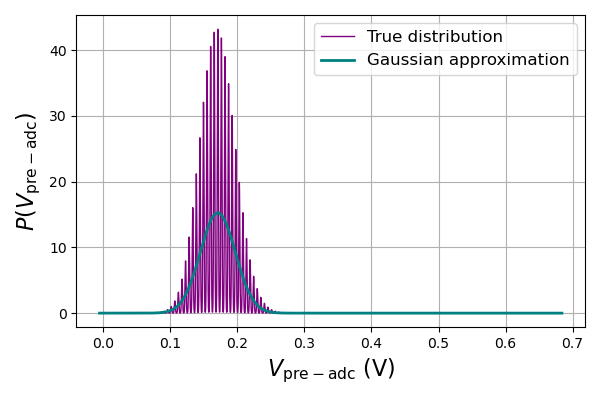}
        \caption{$N = 128$, $\sigma_\mathrm{adc} = \unit[0.75]{mV}$}
    \end{subfigure}  
    \hfill
    \begin{subfigure}{0.32\linewidth}
    \captionsetup{skip=0pt}
        \centering        \includegraphics[width = \linewidth]{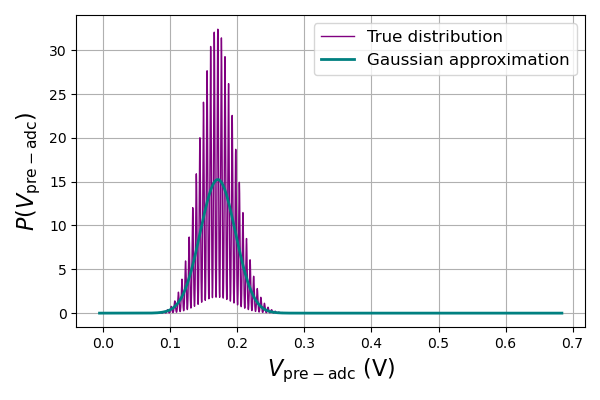}
        \caption{$N = 128$, $\sigma_\mathrm{adc} = \unit[1]{mV}$}
    \end{subfigure}   
    \begin{subfigure}{0.32\linewidth}
    \captionsetup{skip=0pt}
        \centering        \includegraphics[width = \linewidth]{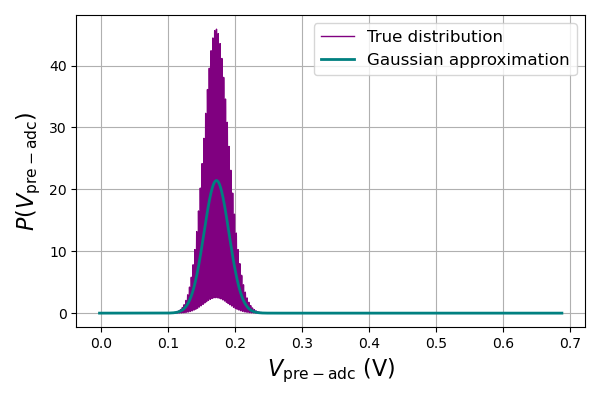}
        \caption{$N = 256$, $\sigma_\mathrm{adc} = \unit[0.5]{mV}$}
    \end{subfigure}
    \hfill
    \begin{subfigure}{0.32\linewidth}
    \captionsetup{skip=0pt}
        \centering        \includegraphics[width = \linewidth]{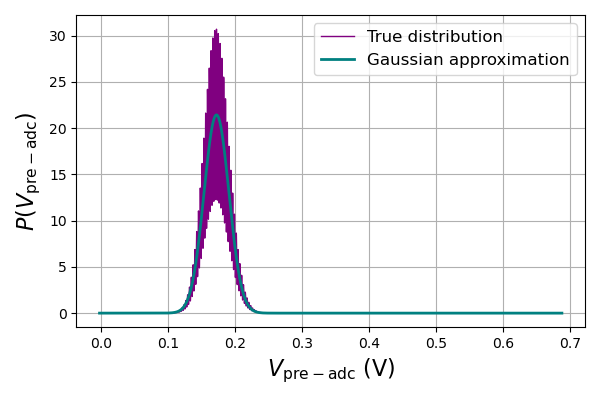}
        \caption{$N = 256$, $\sigma_\mathrm{adc} = \unit[0.75]{mV}$}
    \end{subfigure}  
    \hfill
    \begin{subfigure}{0.32\linewidth}
    \captionsetup{skip=0pt}
        \centering        \includegraphics[width = \linewidth]{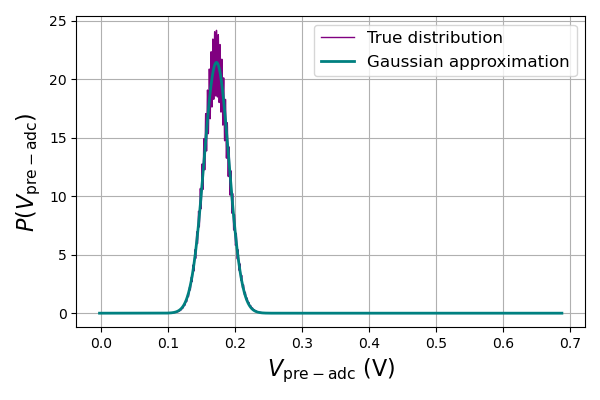}
        \caption{$N = 256$, $\sigma_\mathrm{adc} = \unit[1]{mV}$}
    \end{subfigure}   
    \caption{Impact of dot product dimension ($N$) and standard deviation of ADC thermal noise ($\sigma_\mathrm{adc}$) on the pre-ADC voltage ($V_\mathrm{pre-adc}$) distribution.}
    \label{fig:sim_pre_adc}
\end{figure*}

Figure~\ref{fig:sim_setup} shows the simulation setup used to evaluate the $\mathrm{CSNR}$ of an SRAM-based AIMC. We build a circuit-aware behavioral model of an SRAM-based AIMC chip in a \unit[28]{nm} CMOS process, based on~\eqref{eq:ideal_pmf}–\eqref{eq:pre_adc}, using post-layout simulation data from~\cite{ji2023thesis}. The model computes the value of $\Delta_\mathrm{imc}$ in~\eqref{eq:v_ideal} based on the supply voltage, bit-cell capacitance, and dot product dimension ($N$). It accounts for parasitic capacitance, bit-cell capacitance mismatch and models ADC thermal noise with variance $\sigma_\mathrm{adc}$ as the primary source of the additive Gaussian noise $\eta_\mathrm{a}$ in~\eqref{eq:pre_adc}. We assume $\mathrm{Be}(0.5)$ weights and inputs, \unit[0.9]{V} supply voltage, and \unit[1]{fF} bit-cell capacitance. We present $\mathrm{CSNR}$ results for $\sigma_\mathrm{adc}$ values representative of practical SAR ADC designs~\cite{adc_noise}. We set \(n_\mathrm{samples} = 5 \times 10^5\) and report values of \(\mathrm{CSNR} \leq \unit[50]{dB}\), as this corresponds to  \(>100\) samples producing computational errors. Estimating \(\mathrm{CSNR}\) above \(\unit[50]{dB}\) would require significantly more samples, increasing simulation time. We share a repository containing our behavioral model and CSNR evaluation code: \href{https://github.com/mihirvk2/CSNR-optimal-ADC}{\textcolor{blue}{https://github.com/mihirvk2/CSNR-optimal-ADC}}.


\begin{figure*}[htb]
\setlength{\abovecaptionskip}{0pt} 
    \setlength{\belowcaptionskip}{5pt} 
    \centering    
    \begin{subfigure}{0.32\linewidth}
    \captionsetup{skip=0pt}
        \centering        \includegraphics[width = \linewidth]{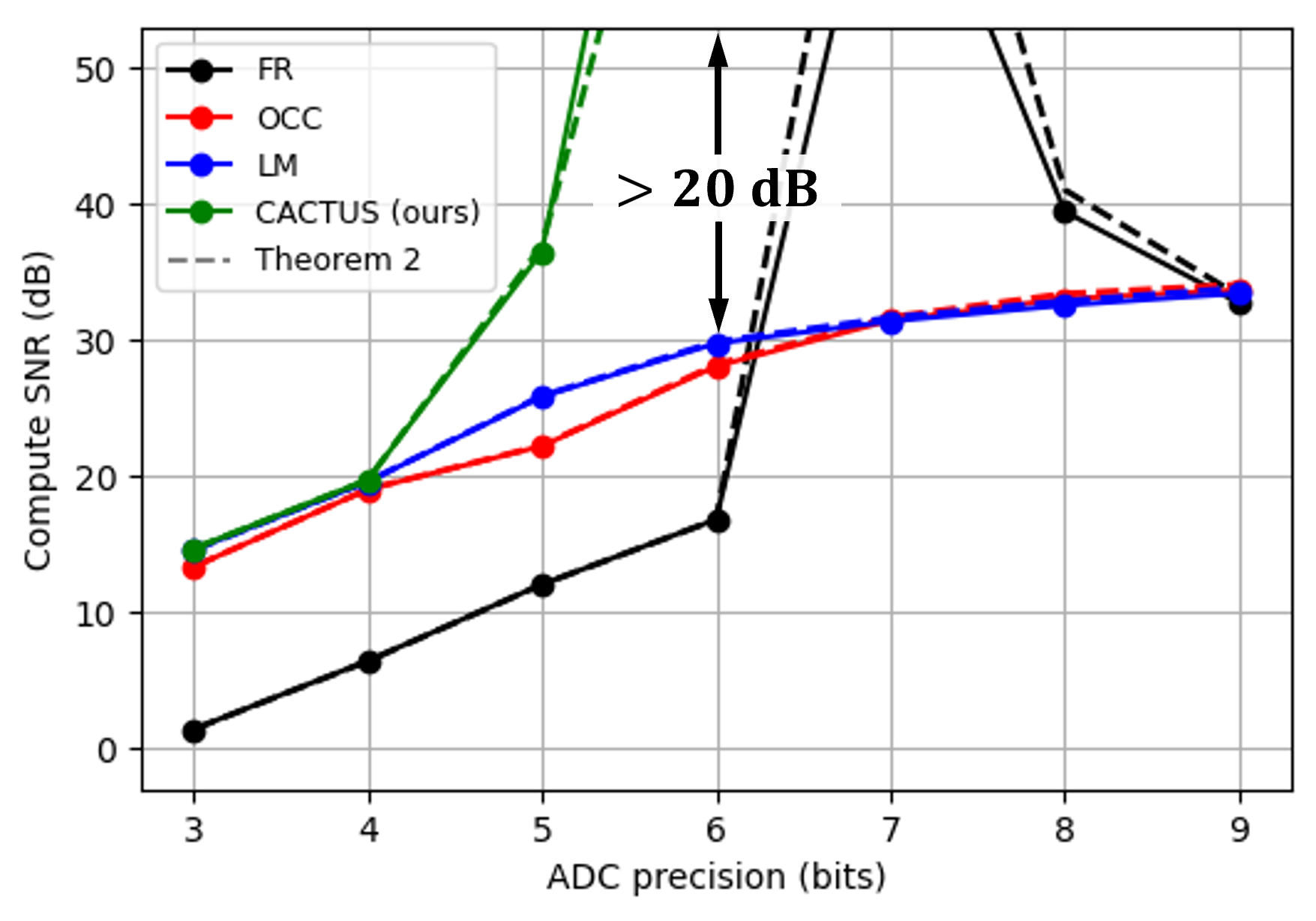}
        \caption{$N = 128$, $\sigma_\mathrm{adc} = \unit[0.5]{mV}$ }
    \end{subfigure}
    \hfill
    \begin{subfigure}{0.32\linewidth}
    \captionsetup{skip=0pt}
        \centering        \includegraphics[width = \linewidth]{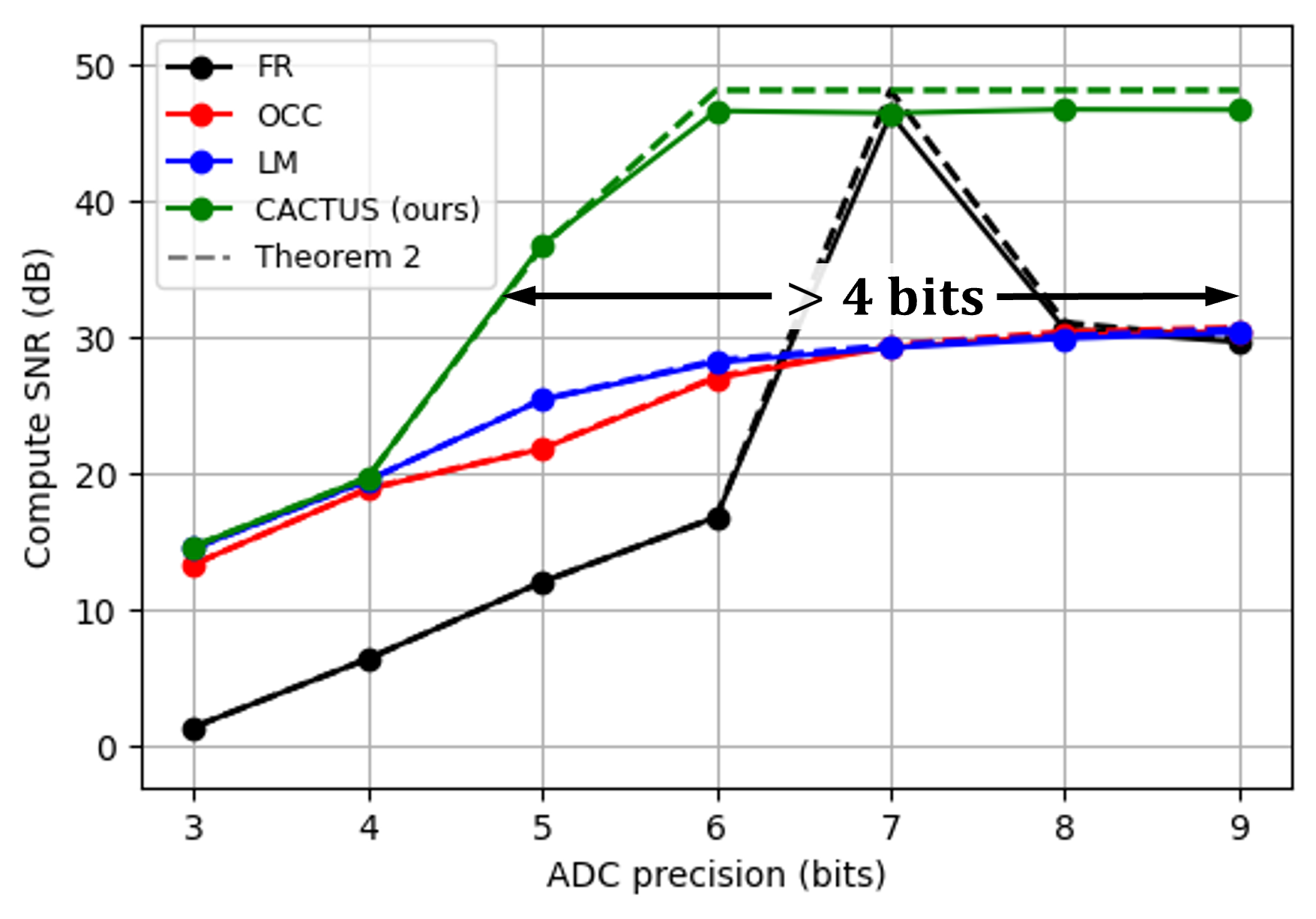}
        \caption{$N = 128$, $\sigma_\mathrm{adc} = \unit[0.75]{mV}$}
    \end{subfigure}  
    \hfill
    \begin{subfigure}{0.32\linewidth}
    \captionsetup{skip=0pt}
        \centering        \includegraphics[width = \linewidth]{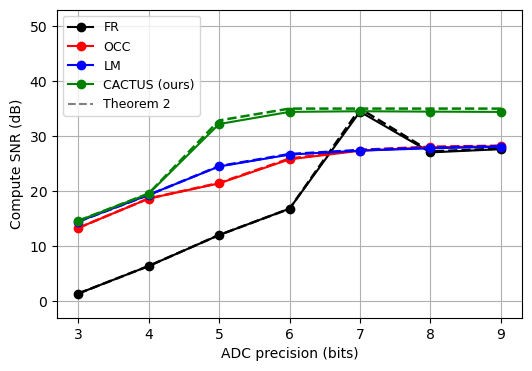}
        \caption{$N = 128$, $\sigma_\mathrm{adc} = \unit[1]{mV}$}
    \end{subfigure}   
    \begin{subfigure}{0.32\linewidth}
    \captionsetup{skip=0pt}
        \centering        \includegraphics[width = \linewidth]{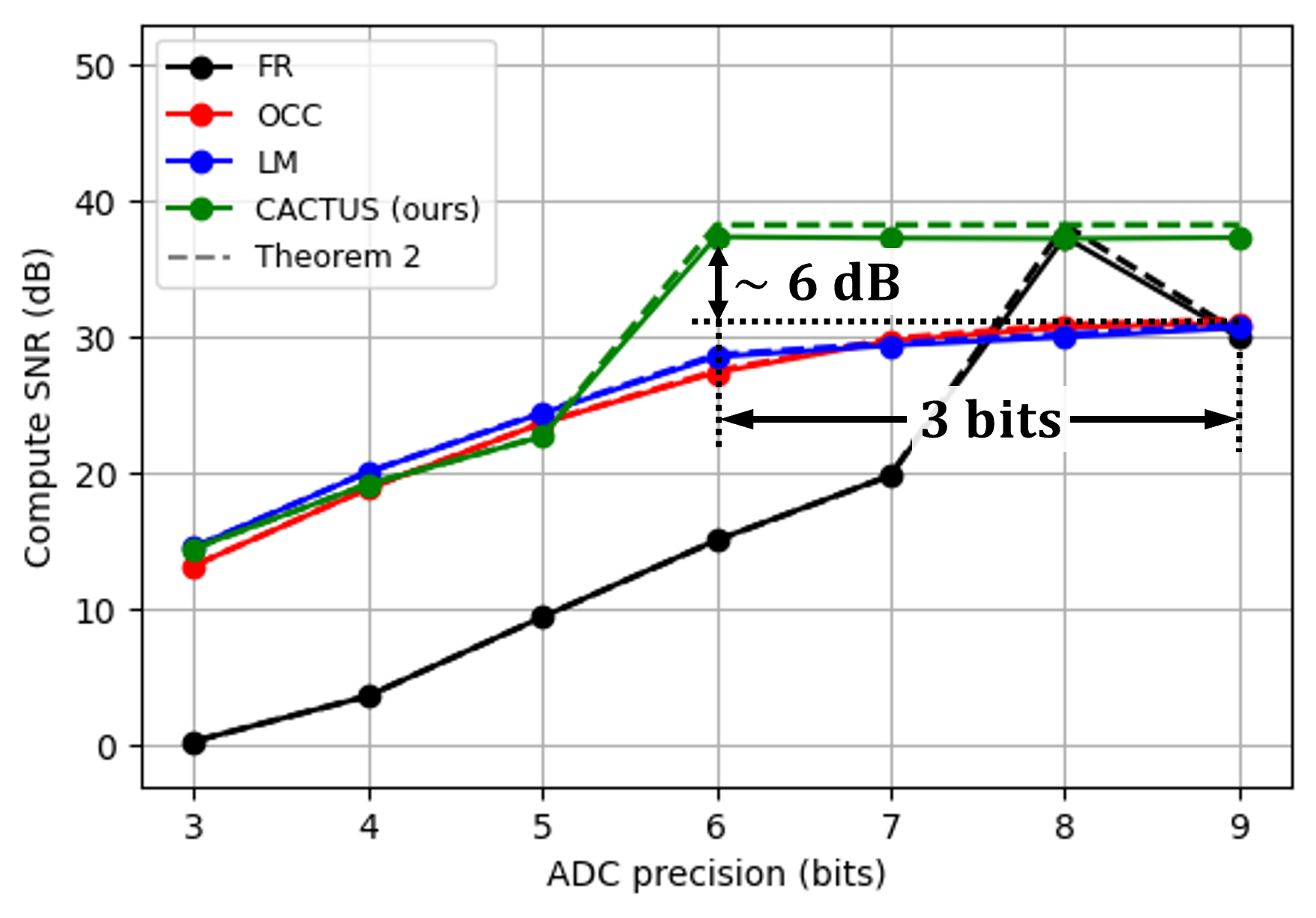}
        \caption{$N = 256$, $\sigma_\mathrm{adc} = \unit[0.5]{mV}$}
    \end{subfigure}
    \hfill
    \begin{subfigure}{0.32\linewidth}
    \captionsetup{skip=0pt}
        \centering        \includegraphics[width = \linewidth]{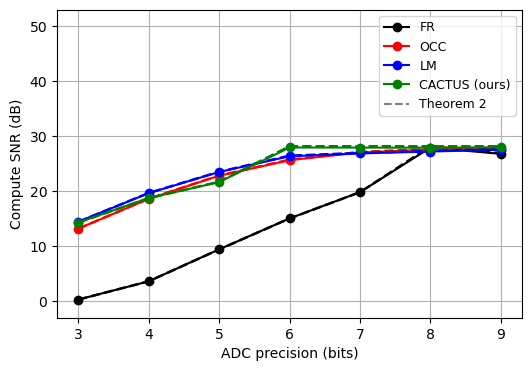}
        \caption{$N = 256$, $\sigma_\mathrm{adc} = \unit[0.75]{mV}$}
    \end{subfigure}  
    \hfill
    \begin{subfigure}{0.32\linewidth}
    \captionsetup{skip=0pt}
        \centering        \includegraphics[width = \linewidth]{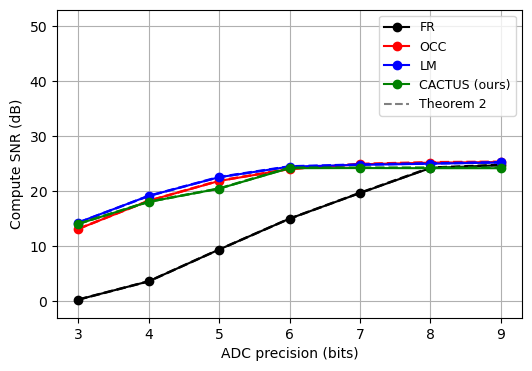}
        \caption{$N = 256$, $\sigma_\mathrm{adc} = \unit[1]{mV}$}
    \end{subfigure}   
    \caption{Impact of ADC quantization scheme on the accuracy of SRAM-based AIMC for different dot product dimensions ($N$) and standard deviations of ADC thermal noise ($\sigma_\mathrm{adc}$).}
    \label{fig:sim_results}
\end{figure*}

\subsection{Impact of $N$ and $\sigma_\mathrm{adc}$ on $V_\mathrm{pre-adc}$}

Figure~\ref{fig:sim_pre_adc} illustrates the impact of dot product dimension ($N$) and ADC thermal noise standard deviation ($\sigma_\mathrm{adc}$) on the pre-ADC voltage distribution ($V_\mathrm{pre-adc}$). Assuming a $\mathrm{Be}(0.5)$ distribution for both inputs and weights, the ideal dot product follows $y_\mathrm{ideal} \sim \mathrm{Bi}(N, 0.25)$. As $N$ increases, the distribution narrows due to the $\sqrt{N}$ scaling of the standard deviation of a binomial random variable. For small $N$ and $\sigma_\mathrm{adc}$, the true distribution deviates significantly from a Gaussian. However, when $N \geq 256$ or $\sigma_\mathrm{adc} \geq \unit[0.75]{mV}$, the resulting Gaussian mixture approaches a Gaussian distribution. As expected and shown later in Fig.~\ref{fig:sim_results}, under these conditions, the gap between $\mathrm{CSNR}$-optimal ADCs and $\mathrm{SQNR}$-optimal ADCs reduces.

\subsection{Impact of $\{\mathbf{t,r}\}$ on $\mathrm{CSNR}$}

 We compare the performance of a CACTUS ADC against the baseline ADCs illustrated in Fig.~\ref{fig:csnr_optimal_adc}:
\begin{enumerate}
    \item \textbf{Full range (FR)}: In an FR uniform ADC, the clipping thresholds in~\eqref{eq:quantizer} are given by: 
    \begin{equation}
        \begin{split}
        \Delta_\mathrm{adc} &= \frac{N \Delta_\mathrm{imc}}{M+1} \\              t_1 = 0.5\Delta_\mathrm{adc} \ &, \ t_M = (M-0.5) \Delta_\mathrm{adc}
        \end{split}
        \label{eq:fr-adc}
    \end{equation}
    \item \textbf{Optimal clipping criterion (OCC)}: In an OCC-based uniform ADC~\cite{sakr2021tsp}, the clipping thresholds are derived based on the assumption that the ADC input follows a Gaussian distribution. Thus
   we approximate the input as a Gaussian and apply OCC.
   \item \textbf{Lloyd-Max (LM)}: The LM algorithm~\cite{lloyd,max} can get trapped in local minima for multi-peak distributions, so we approximate the input as Gaussian for consistent results.
\end{enumerate}

Figure~\ref{fig:sim_results} shows that CACTUS consistently outperforms the FR, LM, and OCC baselines across various operating conditions. We see that CACTUS can provide $>\unit[20]{dB}$ gain in $\mathrm{CSNR}$ at a given ADC precision, reduce the ADC precision requirement by $>\unit[4]{b}$ to meet a target $\mathrm{CSNR}$, and achieve around \unit[6]{dB} $\mathrm{CSNR}$ improvement while simultaneously saving \unit[3]{b} in ADC precision. These benefits diminish as \(N\) and \(\sigma_\mathrm{adc}\) increase, as ADC noise becomes dominant due to a reduced \(\Delta_\mathrm{imc} / \sigma_\mathrm{adc}\) ratio, causing the $\mathrm{CSNR}$- and $\mathrm{SQNR}$-optimal ADC designs to converge. We also observe a $\mathrm{CSNR}$ peak in FR ADC at \(B_\mathrm{adc} = \log_2 N\), where it behaves identically to CACTUS. In this regime, AIMC can achieve a $\mathrm{CSNR}$ significantly higher than the ADC's $\mathrm{SQNR}$, as quantizing \(V_\mathrm{pre-adc}\) to \(V_\mathrm{ideal}\) can cancel the analog noise \((\eta_\mathrm{a})\). This peaking contradicts prior assertions~\cite{gonugondla2020iccad} that $\mathrm{CSNR}$ should increase with $\mathrm{SQNR}$ and is upper bounded by it, but is explained by the theoretical analysis in this paper. 

Figure~\ref{fig:sim_results} also shows that the simulation results closely match the theoretical predictions of Theorem~\ref{thm:main} (dashed lines) and its extensions.



\section{Conclusion} \label{sec:conclusion}
This paper challenges conventional assumptions about the ADC’s role in an AIMC and motivates the exploration of alternative design methodologies that directly optimize for AIMC's $\mathrm{CSNR}$ rather than ADC's $\mathrm{SQNR}$. We propose one such method, based on the theoretical analysis of $\mathrm{CSNR}$-optimal uniform ADCs for AIMCs, and demonstrate its energy and accuracy benefits. The underlying mathematical framework is general enough to support the development of other approaches. More broadly, this work encourages investigation into application-specific ADC designs, where task knowledge can help identify the ADC’s true role at the system level and enable more efficient design optimizations.


\section*{Appendix} \label{appendix}
\noindent \textbf{Proof of Theorem~\ref{thm:offset}}: For a fixed $y$, the probability mass function of $Q_{\mathbf{t,r}}(y \Delta_\mathrm{imc} + \eta_\mathrm{a})$ can be written using~\eqref{eq:quantizer} as follows:
\begin{equation}
    \begin{split}
&\Pr\left(Q_{\mathbf{t,r}}(y \Delta_\mathrm{imc} + \eta_\mathrm{a}) = r_k \right) \\
&= \begin{cases}
    \Phi\left( \dfrac{t_1 - y \Delta_\mathrm{imc}}{\sigma_\mathrm{a}} \right)  \ \ \  \ \ \ \ , \  \text{if } k = 0 \\
    \begin{array}{l}
    \small \Phi\left( \dfrac{t_{k+1} - y \Delta_\mathrm{imc}}{\sigma_\mathrm{a}} \right) 
    - \Phi\left( \dfrac{t_k - y\Delta_\mathrm{imc}}{\sigma_\mathrm{a}} \right)  \\
    \ \ \ \ \ \ \ \ \ \ \ \ \ \ \ \ \ \ \ \ \ \ \ \ \ \ \  , \ \text{if } \  k \in \{1, 2, \dots, M{-}1\}
    \end{array} \\
    1 - \Phi\left( \dfrac{t_M - y \Delta_\mathrm{imc}}{\sigma_\mathrm{a}} \right) \ , \  \text{if }  k = M
\end{cases}
    \end{split}
    \label{eq:imc_quantizer_pmf}
\end{equation}
where $\Phi(\cdot)$ is expressed in~\eqref{eq:normal_cdf}. We use~\eqref{eq:imc_quantizer_pmf} to derive:
\begin{equation}
\begin{split}
&\mathbb{E}_{\eta_\mathrm{a}}\left[Q_{\mathbf{t,r}}\left(y_\mathrm{ideal} \Delta_\mathrm{imc} + \eta_\mathrm{a}\right) \right] \\ &= r_M - \Delta_\mathrm{adc} \sum_{k=1}^{M} \Phi\left( \frac{t_{k} - y_\mathrm{ideal} \Delta_\mathrm{imc}}{\sigma_\mathrm{a}} \right)
\end{split}
    \label{eq:first_moment}
\end{equation}
Using the tower property of conditional expectation, we can write:
\begin{equation}
\begin{split}
    \mathbb{E}\left[y_\mathrm{imc} \right] =\mathbb{E}_{y_\mathrm{ideal}}\left[\mathbb{E}_{\eta_\mathrm{a}}\left[ y_\mathrm{imc} \big| y_\mathrm{ideal}  \right] \right] 
\end{split}
\label{eq:tower1}
\end{equation}
We use~\eqref{eq:first_moment} and \( p_{y_\mathrm{ideal}} \) to evaluate the inner and outer expectations in~\eqref{eq:tower1}, respectively, and substitute the result into~\eqref{eq:zero_mean} to obtain the expression in Theorem~\ref{thm:offset}.

\noindent \textbf{Proof of Theorem~\ref{thm:main}}: We express $\mathrm{MSE_{dp}}$ as,
\begin{equation}
\begin{split}
    \mathbb{E}&\left[\left(y_\mathrm{imc} - y_\mathrm{ideal} \right)^2\right] 
=\frac{\mathbb{E}\left[\left(Q_{\mathbf{t,r}}(y_\mathrm{ideal} \Delta_\mathrm{imc} + \eta_\mathrm{a})\right)^2\right]}{\Delta_\mathrm{imc}^2}  \\
& -\frac{2 \mathbb{E}\left[y_\mathrm{ideal}  Q_{\mathbf{t,r}}(y_\mathrm{ideal} \Delta_\mathrm{imc}
 + \eta_\mathrm{a})\right]}{\Delta_\mathrm{imc}}  +\mathbb{E}\left[y_\mathrm{ideal}^2\right] -\mu_\mathrm{off}^2
\end{split}
\label{eq:expanded_mse}
\end{equation}
\noindent We use~\eqref{eq:imc_quantizer_pmf} to derive: 
\begin{equation}
\begin{split}
\mathbb{E}_{\eta_\mathrm{a}}&\left[\left(Q_{\mathbf{t,r}}\left(y_\mathrm{ideal} \Delta_\mathrm{imc} + \eta_\mathrm{a}\right) \right)^2 \right] \\ &= r_M^2 - 2\Delta_\mathrm{adc} \sum_{k=1}^{M} t_k \Phi\left( \frac{t_{k} - y_\mathrm{ideal} \Delta_\mathrm{imc}}{\sigma_\mathrm{a}} \right)
\end{split}
    \label{eq:second_moment}
\end{equation}
\noindent Using the tower property of conditional expectation, we can write:
\begin{equation}
\begin{split}
    \mathbb{E}\left[\left( y_\mathrm{imc} - y_\mathrm{ideal}\right)^2\right] =\mathbb{E}_{y_\mathrm{ideal}}\left[\mathbb{E}_{\eta_\mathrm{a}}\left[ \left(y_\mathrm{imc} - y_\mathrm{ideal}\right)^2 \big| y_\mathrm{ideal}  \right] \right] 
\end{split}
\label{eq:tower2}
\end{equation}
We use~\eqref{eq:first_moment},~\eqref{eq:second_moment}, and \( p_{y_\mathrm{ideal}} \) to evaluate the inner and outer expectations in~\eqref{eq:tower2}, respectively, and substitute the result into~\eqref{eq:expanded_mse} to obtain the expression in Theorem~\ref{thm:main}.

\bibliographystyle{IEEEtran}
\bibliography{references}

\end{document}